\documentclass[%
 aip,reprint, twocolumn,nofootinbib 
]{revtex4-1}

\usepackage{amsmath,bm,graphicx}
\usepackage{amssymb}
\usepackage{amsfonts}
\usepackage{graphicx}
\usepackage{xcolor}
\usepackage{epstopdf}
\usepackage{mathrsfs}
\usepackage{latexsym}
\usepackage{hyperref}
\usepackage{wasysym}
\usepackage{dcolumn}
\usepackage{bm}
\usepackage{natbib}
\usepackage{float}
\usepackage{cancel}
\usepackage{comment}
\usepackage{soul}
\usepackage[normalem]{ulem}

\newcommand{\pa}[1]{\textcolor{blue}{#1}}

\newcommand{\nn}{\nonumber\\}
\newcommand{\eqr}[1]{Eq.~\eqref{#1}}
\newcommand{\upd}{\mathrm{d}}

\newcommand{\LL}{{\mathcal{L}_{13}}}
\newcommand{\LLo}{{\mathcal{L}_{13}^o}}
\newcommand{\TT}{{\mathcal{L}_{33}}}
\newcommand{\TTo}{{\mathcal{L}_{33}^o}}

\makeatletter
\def\@email#1#2{%
 \endgroup
 \patchcmd{\titleblock@produce}
  {\frontmatter@RRAPformat}
  {\frontmatter@RRAPformat{\produce@RRAP{*#1\href{mailto:#2}{#2}}}\frontmatter@RRAPformat}
  {}{}
}%
\makeatother

\begin{document}

\title{Transport of electrolytes across nanochannels: the role of  slip}

\author{M. Florencia Carusela}
\affiliation{Instituto de Ciencias, Universidad Nacional de General Sarmiento, Los Polvorines, Buenos Aires, Argentina}
\affiliation{National Scientific and Technical
Research Council, Argentina}
\author{Jens Harting}
\affiliation{Helmholtz Institut Erlangen-N\"urnberg  for Renewable Energy (IET-2), Forschungszentrum J\"ulich, Cauerstr. 1, 91052, Erlangen, Germany}
\affiliation{Department of Chemical and Biological Engineering and Department of Physics, Friedrich-Alexander-Universit\"at Erlangen-N\"urnberg, Cauerstr.\,1, D-91058 Erlangen, Germany}
\author{Paolo Malgaretti}
\affiliation{Helmholtz Institut Erlangen-N\"urnberg  for Renewable Energy (IET-2), Forschungszentrum J\"ulich, Cauerstr. 1, 91052, Erlangen, Germany}

\email{p.malgaretti@fz-juelich.de}
\email{fcarusela@campus.ungs.edu.ar}
\begin{abstract}

We characterize the electrokinetic flow due to the transport of electrolytes embedded in nanochannels of varying cross-section with inhomogeneous slip on their walls, modeled as an effective slip length on the channel wall. We show that, within linear response and Debye-H\"uckel regime, the transport coefficients, and so the fluxes, can be significantly improved by the presence of a hydrophobic surface coating located at the narrowest section of the nanochannel.
Our model indicates that the enhancement is larger when considering electric conductive walls in comparison to dielectric microchannel walls, and it is produced by a synergy between the entropic effects due to the geometry and the presence of the slip boundary layer. Our results show that a tailored hydrophobic coating design can be an effective strategy to improve transport properties in the broad areas of lab-on-a-chip, biophysics, and blue energy harvesting and energy conversion technologies.
\end{abstract}

\maketitle

\section{Introduction}\label{sec:Introduction}

Electrolytes play a key role in a wide range of research areas from biological fluids \cite{Ghader2020, Lapizco2021,Reddy2022,Gross1968}, ion channels and membranes \cite{chen2018, Zhou2020, Sun2023, Aluru2023,Green2022,Peters2016} to energy storage and conversion cells \cite{Guth2009,Chouhan2011,Wu2022,Ibrahim2021,LiLi2024,DeBiswajit2020,SHI2022,cheng2022}. Advances in microfabrication technologies have broadened the spectrum of electrolyte-based applications, largely due to the development of micro- and nanofluidic-based devices, which have many advantages over their macrofluidic counterparts \cite{nguyen2019fundamentals}. These devices have the unique property of a relatively large surface to volume ratio, which allows for a high ability to control flows and species at selected locations with microscopic precision \cite{Sparreboom2009, Nasiri2020}, improving mass and energy transfer \cite{chen2018, Han2020}. In addition, microfluidic systems also offer many practical benefits including reduced manufacturing costs, shorter analysis times and integration on lab-on-chip platforms \cite{mi9090461}. Fluidic channel systems typically operate under the influence of pressure, voltage, temperature, or salinity gradients \cite{Werk2020, Sun2023}. However, when channels have characteristic dimensions in the micro/nanometre range, the influence of surface effects, capillarity, wetting and electrical double layer formation can affect the transport properties \cite{Wang2020,Mahapatra22,Zeng2024, Ge2020, Adjary2006}. The electrical double layer \cite{Mesliyah2006, Novotny2017} adjacent to the  walls can induce electrokinetic flows that can affect the global and local transport properties of the system. This layer is formed by the contact of the electrolyte solution with the microchannel walls, creating a surface charge with the consequent redistribution of free ions. When an external electric field, pressure or chemical difference is applied to the ends of the nanochannel, electrosmotic flow occurs.
\begin{figure}[t]
\centering
\includegraphics[width=0.4\textwidth]{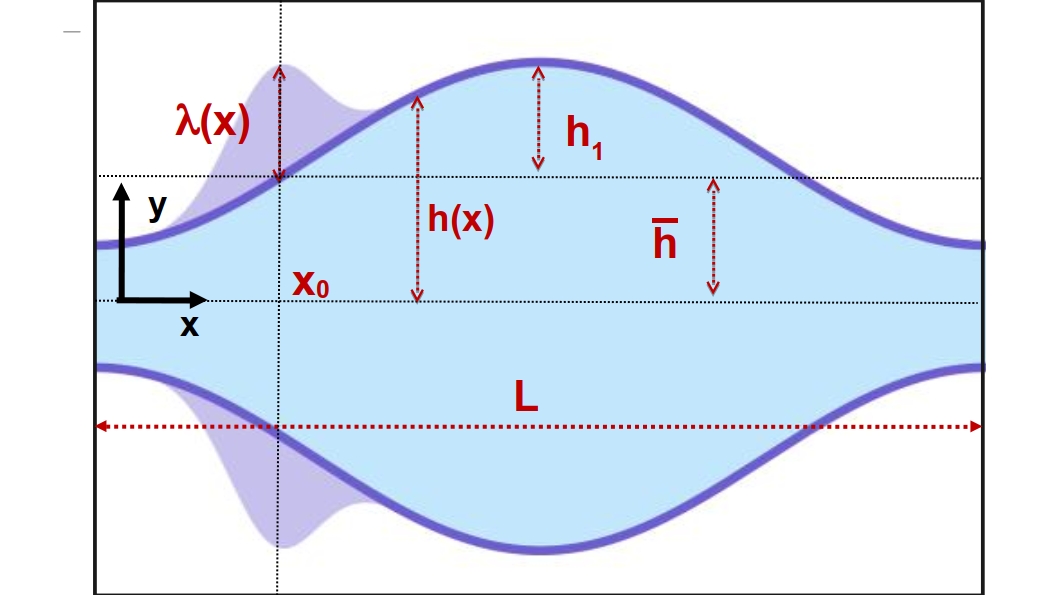}
\label{sketch}
\caption{Schematic view of the system. The light blue region indicates a microchannel of length $L$, average cross section $\bar{h}$ and corrugation $h_1$. Dark violet lines represent the physical walls and light purple regions indicate the "fictious" slip length centered at an arbitrary position $x_0$.}
\end{figure}

A significant challenge in modelling these systems is posed by the description of the fluid/wall interactions, particularly the hydrophobic nature of the surface. These characteristics define the boundary conditions for the flow, which is a crucial point to consider when fluids flow in micro or nano-confinements. The assumption that the relative velocity of the fluid with respect to the wall is zero (no-slip boundary condition) is usually assumed to be valid because it is in good agreement with macroscopic observations. However the emergence of new techniques with micro or nanoscale resolution produced new experimental evidence of a reduced resistance to liquid flow at nonwettable solid walls, which has been theoretically and numerically explained as the existence of a fluid slip \cite{neto2005,Wu2017}. A fluid slip means that the tangential velocity of the fluid differs from that of the solid wall and can be quantified by an "effective" distance, or slip length, beyond the solid wall at which the fluid velocity is assumed to be zero. It should be understood as the equivalent slip required on a smooth uniform solid surface that would produce the same flow conditions far away from the surface. The occurrence of slip implies a significantly reduction of the hydrodynamic resistance of the fluid to flow, an effect that becomes relevant at microscopic scales. The concept of effective slip length can be extremely useful to describe many experiments and practical applications \cite{Kunert2010,Schmiescheck2012,Crick23,Rasitha2024}.
In this scenario, the optimisation of the transport properties, and thus the fluxes (of mass, charge, and heat) along micro and nanochannels, has become crucial in the design of suitable microfluidic platforms for incorporation into biological, energy- or thermal-based electrochemical devices \cite{FERNANDEZLAVILLA2019175}. To address this challenge 
theoretical modelling and numerical studies are necessary to understand the impact of the base fluid, geometry and materials of the channel, as well as the interaction between the walls and the fluid (hydrophobic slippage) on the electrokinetic phenomena \cite{PM19,Vinogradova2021,Vinogradova2022,Vinogradova2023,CHO201294,BANERJEE201917,Joly2004}.

Many theoretical and numerical studies about the role of hydrophobic slippage on mass, charge and heat transport consider pores or microchannels of constant sections \cite{Harting2010,Zhang2022,Ranjan2023}. 
However, many technological applications rely on transport across porous materials within which the constant section of the conduit is not fulfilled.  Accordingly, in order to optimize the transport properties of these devices it would be desirable to establish a synergic engineering of microchannel geometry  and surface coatings (wettability and slippage). 
In fact, it has been demonstrated that entropic barriers, as observed in static \cite{Zwanzig, entropic2, entropic3,MPR2013} or time-dependent~\cite{CR2017,Carusela2021} varying cross-section microchannel, can be employed to enhance transport, as well as to develop  rectifiers \cite{Lairez2016,Malgaretti2014EPJ,CR2018}, and particle "splitters"~\cite{reguera2012,motz2014}, which have a multitude of beneficial micro and nano-technological applications.  

Here, we present a novel model within which both the local slip length and the geometry of the microchannel are accounted for. 
The Onsager matrix is derived in the presence of a hydrophobic slippage on the basis of the lubrication approximation, which relies on the length scale separation between longitudinal and transverse characteristic lengths.
On the top of this, the Poisson-Boltzmann equation has been linearized (Debye-H\"uckel regime) buy assuming low ionic concentrations and $z-z$ electrolytes. In particular, we focus on the role of the "direct" electrostatic interactions and hence we disregard van der Waals interactions which may lead to diffusio-osmotic fluxes (as reported in  Ref.~\cite{Adjary2006}).
The electric, solute and solvent fluxes, which in the linear response regime are proportional to the applied electric fields, ion concentrations, and pressure differences, are analyzed for both conducting and dielectric nanochannel walls. 

The structure of the text is as follows. In Section 2, we introduce our model and the set of electrokinetic equations used to describe solute and solvent fluxes. In Section 3, we present the transport equations and the calculation of the Onsager matrix. In Section 4, we discuss scenarios for transport improvement. Finally, this article concludes with Section 5.

\section{The model}\label{sec:Model}

We analyze a varying cross section nanochannel filled with a $z-z$ electrolyte embedded in a solvent of dielectric constant $\epsilon$. The solvent is incompressible and has a viscosity $\eta$. The channel has a length $L$ and it is translationally invariant along the $x$-direction with a varying channel width 
\begin{align}\label{eq:hx}
 h(x)=\bar{h} -h_1 \cos(2 \pi x /L) 
\end{align}
in the $y$-direction, where $\bar{h}=(1/L) \int_0^L h(x)dx$ is the average width, and $h_1$ is the amplitude of the corrugation. 
On the physical walls there is a hydrophobic region where the fluid partially slips on the channel walls.
We consider that the system can be driven out of equilibrium by three thermodynamic forces given by:
a) a difference of eletrochemical potentials between $x=0$ and $x=L$, due to the contact with two chemostats at electrochemical potentials $\mu^\pm(x=0)$ and $\mu^\pm(x=L)$, respectively; 
b) a potential difference $\Delta V$; 
c) a pressure difference $\Delta P$ between $x=0$ and $x=L$.
As we consider isothermal systems at temperature $T$, there is no force due to a temperature gradient, therefore local heat generation can be neglected. Besides, our model does not account for surface conduction in the Stern layer.
      
The steady state of our system can be modeled by a set of classical electrokinetic equations obtained from the combination of the Poisson, Nernst-Planck and Stokes equations \cite{PM19,Andelman_book,SoftMat}:
\begin{subequations}\label{eq:electrokinetic}
\begin{align}
\epsilon\nabla^{2}\psi(x,y)&=-zeq(x,y)\,,\label{eq:poisson}\\
 \textbf{j}^{\pm}(x,y)&=\rho^{\pm}(x,y)\left[\textbf{v}(x,y)-D\beta \boldsymbol{\nabla}\mu^{\pm}(x,y)\right]\,,\label{eq:nernstplanck}\\
 \boldsymbol{\nabla}\cdot\textbf{j}^{\pm}(x,y)&=0\,,\label{eq:continuity}\\
 \eta\nabla^2 \textbf{v}(x,y)&=-\textbf{F}_{\text{tot}}(x,y)+\boldsymbol{\nabla}P_{\text{tot}}(x)\,,\label{eq:stokes}\\
 \boldsymbol{\nabla}\cdot\textbf{v}(x,y)&=0\,.\label{eq:incompressibility}
\end{align}
\end{subequations}
$\psi(x,y)$ is the electrostatic potential inside the nanochannel, which is determined by the Poisson equation \eqref{eq:poisson} and $\textbf{v}$ the flow velocity. $D$ is the diffusivity of the electrolyte, $z$ the 
valence, $e$ the elementary charge, and $\beta=1/k_B T$ the inverse of the thermal energy where $k_B$ is the Boltzmann constant, and 
$T$ the absolute temperature.
The local charge number density
\begin{align}\label{eq:def-q}
q(x,y)=\rho^{+}(x,y)-\rho^{-}(x,y)
\end{align}
is obtained from the difference between local cationic and anionic number densities, 
$\textbf{j}^{\pm}$ are the ionic currents due to advection and diffusion. These currents are modeled by the Nernst-Planck equation \eqref{eq:nernstplanck} describing the dynamics of point-like ions which represents a good approximation for dilute electrolytes in small electric fields. 

If the system is driven by a pressure drop, and considering a regime characterized by a low Reynolds number, the electrolyte solution will flow according to the Stokes equation \eqref{eq:stokes}, 
with $\partial_xP_{\text{tot}}(x)=\partial_xP(x)+\Delta P/L$, where $\partial_x P$ is the $x$-component of the geometrically-induced local pressure gradient that is determined by the boundary conditions and by fluid incompressibility. In addition, 
the $x$-component of the total electrostatic force density acting on the fluid is given by 
\begin{equation}
F_{\text{tot}}(x,y)=-zeq(x,y)\partial_x \psi(x,y)\,.
\label{eq:def-tot-E0}
\end{equation}
Finally, equations considering steady-state continuity and fluid incompressibility are given by 
Eqs.~\eqref{eq:stokes} and \eqref{eq:incompressibility}. 
Equations~\eqref{eq:electrokinetic} are solved according to the following boundary conditions
\begin{subequations}\label{eq:electrokineticbc}
\begin{align}
&\mathbf{n}^\pm\cdot\mathbf{j}^{\pm}(x, \pm h(x))=0\,,\label{eq:noflux}\\
&\textbf{v}(x, \pm H(x))=0\,,\label{eq:noslip}
\\
&\psi\to
\begin{cases}
\mathbf{n}^\pm\cdot \nabla\psi(x,y)|_{y=\pm h(x)}&=\mp \frac{e\sigma}{\epsilon} \quad\text{diel.}\\
\psi(x,\pm h(x))&=\zeta \quad\quad\text{cond.}
\end{cases}
\end{align}
\end{subequations}
where $\mathbf{n}^\pm$ is the local normal at the channel walls, $\sigma$ is the surface charge density, $\epsilon$ is the dielectric constant of the solvent and $\zeta$ is the surface potential.
Equations \eqref{eq:noflux} and \eqref{eq:noslip} states that the no-flux and no-slip boundary conditions, respectively.
Although hydrophobic coatings can have an effect on surface charge density, this effect may be very small depending on the characteristics of the coatings, the type of electrolyte and the material and surface of the nanochannel. In order to focus on the role of the slip length on the transport, we consider a homogeneous electrostatic boundary condition which is a reasonably good approximation when these factors have a negligible effect. An extension to the case with a non constant surface charge density can be developed within this framework, however this goes beyond the central point of the present work.
In addition, we implement boundary conditions on $\psi(x,y)$ that depend on the conductive properties of the channel walls. We impose a constant $\zeta$ potential for conductive walls and a constant surface charge $\sigma$ for the dielectric ones. We denote the electrostatic potential as  $\psi^{\zeta}(x,y)$ or $\psi^{\sigma}(x,y)$, respectively.

In order to get analytical insight, we restrict ourselves to nanochannels whose variation in the section, $\Delta h$, occurs on long length scales, namely $\frac{\Delta h}{L}\ll 1$\footnote{Typically $\Delta h/L \lesssim 0.1$ is enough to be within the lubrication approximation.}. In fact, for $\frac{\Delta h}{L}\ll 1$ changes of $v_x$ along the longitudinal direction are much smaller than those along the transverse direction. For what concerns the electrostatic field, $\psi$, the length scale separation should be between the screening length and the channel width. As we will specify later, our model relies on the  linearized Poisson-Boltzman equation (Debye-H\"uckel approximation), and hence the lubrication approximation requires the Debye length, $k_0^{-1}$, to be much smaller than the channel length, $L$. Since the typical range of Debye length is $k_0^{-1}\lesssim 100$nm, the length scale separation $k_0^{-1}\ll L$ is typically fulfilled in microchannels whose lengths exceed $\simeq 100$nm. This means that due to this longitudinal and transverse length scale separation (fast equilibration in the transversal direction) $\partial_{x}^{2}$ terms in Eqs.~(\ref{eq:poisson}) and (\ref{eq:stokes})  become negligible as compared to $\partial_{y}^{2}$ terms.  Doing this ``lubrication-like'' approximation \eqr{eq:poisson} and \eqr{eq:stokes} become analytically solvable.
Within the lubrication approximation the partial slip on the walls is captured by the  
Navier slip condition
\begin{align}
    v|_{y=\pm h(x)} = \ell(x) \partial_y v|_{y=\pm h(x)}\,,
    \label{eq:navier-BC}
\end{align}
where $\ell(x)$ is the slip length. 
Accordingly, the Navier boundary conditions lead to the velocity profile
\begin{align}
    v_\ell(x,y) = -\frac{\nabla P(x)}{\eta}\left(y^2-h^2(x)-2\ell(x) h(x)\right)\,.
    \label{eq:navier}
\end{align}
Eq.~\eqref{eq:navier} is equivalent (see Appendix~\ref{app:A}) to imposing no slip condition on a fictitious boundary
\begin{equation}\label{eq:def-lambda}
H(x)=h(x)+\lambda(x)    
\end{equation}
with 
\begin{align}
    \lambda(x) = h(x)\left(\sqrt{1+2\frac{\ell(x)}{h(x)}}-1\right)\,,
\end{align}
which we find more convenient to use.

We consider that the electrolyte is dilute so that we can linearize the Poisson-Botlzmann equation and reduce it to the Debye-H\"uckel for which analytical expressions can be obtained\footnote{The Debye-H\"uckel approximation is valid for $z-z$ electrolytes and for weak electrostatic potential at the walls. See Ref.~\cite{Andelman_book} for a more detailed discussion.}. For conducting channel walls the equilibrium electrostatic potential reads
\begin{equation}
\psi^\zeta_0(x,y)=\zeta\frac{\cosh\left[k_0y\right]}{\cosh\left[k_0h(x)\right]}\,,
 \label{eq:electric_potential-cond}
\end{equation}
while for dielectric walls it reads
\begin{equation}
\psi^\sigma_0(x,y)=\frac{e\sigma}{\epsilon k_0}\frac{\cosh\left[k_0y\right]}{\sinh\left[k_0h(x)\right]}\,,
\label{eq:electric_potential-0-1}
\end{equation}
where $k_0=\sqrt{\beta (ze)^2 \gamma_0/\epsilon}$ is the inverse Debye length and $\gamma_0=2\varrho_0$ is the salt number density \cite{PM19}.

As we mentioned above, we consider that the system can be driven out of equilibrium by applied external forces given by $\Delta P/L$, $ze\Delta V/L$, and $\Delta \mu/L$ and we focus on the liner response regime where the fluxes are linear in the forces.
Following a similar strategy as  developed in Ref.~\cite{PM19}, we can obtain, at leading order in lubrication, the solution $v_x(x,y)$ of the Stokes equation [Eq.~\eqref{eq:stokes}] subjected to the no-slip boundary condition on $H(x)$ [Eq.~\eqref{eq:lambda}]. 
As we mentioned, $\lambda(x)$ mimics the effective slip length: $\lambda(x) = 0$ corresponds to the no-slip case, whereas $\lambda(x)\neq 0 $ gives rise to a finite slip length. 
To do this, we split the velocity $v_x(x,y)$ into an electro-osmotic contribution $u_{\textrm{eo}}$ due to the drag of ions along the fluid caused by the presence of an electric field, and a pressure-driven contribution $u_{P}$
\begin{subequations}\label{eq:vel_field_x-y}
\begin{align}
v_x(x,y)&= u_{P}(x,y)+u_{\textrm{eo}}(x,y)\,,\\
u_{P}(x,y)&=\frac{\partial_xP_{\text{tot}}(x)}{2\eta}\left[y^{2}-H^{2}(x)\right]\,,\label{eq:vxpressure}
 \end{align}
\end{subequations}
The local pressure $\partial_x P_\text{tot}(x)=\partial_xP(x)+\Delta P/L$,  
accounts for the applied pressure drop $\Delta P$ between $x=0$ and $x=L$, as well as for the local pressure $P(x)$ that ensures fluid incompressibility [\eqr{eq:incompressibility}]. 
Inserting \eqr{eq:vel_field_x-y} into the volumetric fluid flow leads to
\begin{align}
 Q &= \int_{-h(x)}^{h(x)} v_x(x,y)\,\upd y\,\nonumber\\
 &= \int_{-h(x)}^{h(x)} u_P(x,y)\,\upd y+\int_{-h(x)}^{h(x)} u_{eo}(x,y)\,\upd y\,.
 \label{eq:Q}
 \end{align}
Performing the $y$-integral over $u_P$, integrating the expression over $\int_{0}^{L}\upd x$ , imposing fluid incompressibility ($\partial_x Q=0$), and using $\int_0^L \partial_x P_\text{tot}(x) \,\upd x =\Delta P$ \cite{PM19}
the fluid flow can be expressed as the sum of two different contributions:
\begin{subequations} \label{eq:Q-lin}
\begin{align}
 Q&\equiv Q_P+ Q_{\textrm{eo}}\,,\\
 Q_P&=-\frac{1}{3H_{3}}\frac{ \bar{h}^3 \Delta P}{\eta L}\,,\label{eq:QP}\\
 Q_{\textrm{eo}}&=\frac{\bar{h}^3}{H_{3}L}\int\limits_0^L\frac{\upd x}{h(x)(3 H^2(x)-h^2(x))}\int\limits_{-h(x)}^{h(x)}\!\!u_{\textrm{eo}}(x,y)\,\upd y\label{eq:QEO}\,.
\end{align}
\end{subequations}
Here $Q_{P}$ is the pressure-driven volumetric fluid flow and $Q_{\textrm{eo}}$ the electroosmotic flow and $H_{3}$
\begin{equation}\label{eq:H3}
H_{3}\equiv \frac{\bar{h}^{3}}{L}\int_{0}^{L}\!\!\frac{1}{h(x) (3 H^2(x)-h^{2}(x))}\,\upd x
\end{equation}
is a position independent coefficient that fulfills $H_{3}\ge 1$ and $\bar{h}$ the average channel section. 

Finally, in order to determine $Q_{\rm eo}$ we need to characterize the ionic transport. 
For weak external forces, within the Debye-H\"uckel regime and at first order in lubrication, we expand the non-equilibrium electric potential, charge densities and electrochemical potential about their equilibrium values (small-force $f$ expansion):
\begin{subequations}
\begin{align}
\psi(x,y)&=\psi_0(x,y)+\psi_{0,f}(x,y)+\mathcal{O}(f^2)+\mathcal{O}(\psi_0^3)\,,\label{eq:def-phi-exp-f}\\
 \rho^{\pm}(x,y) &=  \rho_{0}^{\pm}(x,y)+\rho^{\pm}_{0,f}(x,y)+\mathcal{O}(f^2)+\mathcal{O}(\psi_0^2).
\label{eq:def-rho-exp-f}
\end{align}
\end{subequations}
Hence, this expansion for small values of $f$ about the Debye-H\"uckel solution is meaningful provided that contributions of order $\mathcal{O}(f)$ are larger than those of order $\mathcal{O}(\psi_0^3)$. 
From hereon, and in order to simplify the notation, in all $\mathcal{O}(f)$ terms that we write, we refer to the lubrication approximation,
in particular $\rho^{\pm}_{0,f}\to \rho^{\pm}_{f}$ and $\psi_{0,f}\to \psi_{f}$.
From these expressions
we find an expansion of the chemical potential,
 $\mu^\pm(x,y)=\mu^\pm_0+ \mu^\pm_f(x,y)+\mathcal{O}(f^2)$
with 
\begin{align}\label{eq:electrochemicalpotential}
 \beta \mu^\pm_f(x,y)&=\frac{\rho^\pm_f(x,y)}{\rho^\pm_0(x,y)}\pm \beta ze \psi_f(x,y)\,.
\end{align}
Assuming a small transverse Peclet number, that is $\bar{h} v_{y}/D\ll1$, the steady state is achieved by systems that are in \textit{local equilibrium} $\partial_{y}\mu(x,y)=0$ in every transverse section of the microchannel located at $x$. Therefore, 
at linear order in $\psi_0$, 
the density profiles can be expressed as
\begin{align}
\rho^{\pm}_{f}(x,y)&=   \varrho_{0}\left[\beta \bar{\mu}^{\pm}_f(x)\mp\beta ze \psi_f(x,y)\right]\left[1\mp\beta ze\psi_0(x,y)\right]\,.\label{eq:def-rho-exp-f2}
\end{align}
where $\mu^\pm_f$ defines the intrinsic electro-chemical potential as $ \bar{\mu}^{\pm}_f(x)\equiv \mu^\pm_f(x,y)$.

\section{Transport Equations}\label{sec:TransportEq.}

The next step is to establish the corresponding transport equations for the system.
The steady-state continuity \mbox{equation \eqref{eq:continuity}}, together with the no-flux boundary condition 
\eqr{eq:noflux}, implies the $x$-independence of the following cross-sectional integrals 
\begin{align}
J^{\pm}&=\int_{-h(x)}^{h(x)}\! j_{x}^{\pm}(x,y) \,\upd y\,, 
\label{eq:coupled_fokker_planck}
\end{align}
which represent the total ionic fluxes through a slab at $x$.
Following \cite{PM19}, the expressions for the solute $J_{c}=J^{+}+J^{-}$ 
and charge $J_{q}=J^{+}-J^{-}$ fluxes are obtained by inserting Eqs.~\eqref{eq:nernstplanck} and \eqref{eq:def-rho-exp-f2} into \eqr{eq:coupled_fokker_planck} obtaining
\begin{subequations}\label{eq:lmbd-eps-int}
\begin{align}
\frac{J_{c}}{D}&=\frac{ \gamma_{0}Q}{D}+\beta ze\overline{\psi_0}(x)\partial_x \xi_{f}(x)-2h(x)\partial_x \gamma_{f}(x)\nn
&\quad+\mathcal{O}(f^2)\,,\label{eq:lmbd-eps-int1}\\
\frac{J_{q}}{D}&=\frac{\mathcal{J}_{q} (x)}{D}+\beta ze\overline{\psi_0}(x)\partial_x \gamma_{f}(x)-2h(x)\partial_x \xi_{f}(x)\nn
&\quad+\mathcal{O}(f^2)\,,\label{eq:lmbd-eps-int2}
\end{align}
\end{subequations}
where $  \gamma_{f}(x),   \xi_{f}(x), \overline{\psi_0}(x)$, and $\mathcal{J}_{q} (x)$ are defined as
\begin{subequations}
\begin{align}
  \gamma_{f}(x)&= \varrho_{0}\beta \left[\bar\mu^+_f(x)+\bar\mu^-_f(x)\right]\,,\label{eq:def-varphi}\\
  \xi_{f}(x)&= \varrho_{0}\beta\left[ \bar\mu^+_f(x)- \bar\mu^-_f(x)\right]\,,\label{eq:barmupm}\\
\overline{\psi_0}(x)&\equiv\int_{-h(x)}^{h(x)}\! \psi_{0}(x,y) \,\upd y\,, \\
 \mathcal{J}_{q} (x)&\equiv\int_{-h(x)}^{h(x)}q_0(x,y) v_{x}(x,y)\,, \label{eq:Q-Q}\\
 &\equiv \int\limits_{-(x)}^{h(x)}q_0(x,y) u_P(x,y)+\int\limits_{-h(x)}^{h(x)}q_0(x,y) u_{eo}(x,y)  \,\upd y\,.\nonumber 
\end{align}
\end{subequations}
As we consider expansions up to $O(\psi_0)$, the second term on the right hand side of Eq.~\eqref{eq:Q-Q} can be neglected. We remark that corrections to $\mathcal{J}_{q}$ due to the electroosmotic flow may arise when the full Poisson-Boltzmann (and not the linearized Debye-H\"uckel) equation is accounted for.  Using that $q(x,y)=\rho^{+}(x,y)-\rho^{-}(x,y)$ [Eq.~\eqref{eq:def-q}] and Eq.\eqref{eq:vxpressure}, $\mathcal{J}_q$ can be rewritten as
\begin{subequations}
\begin{align}  
  \mathcal{J}_{q} (x)&\equiv \int\limits_{-h(x)}^{h(x)}(\rho^{+}(x,y)-\rho^{-}(x,y)) \frac{\partial_{x}P}{2 \eta}[y^2-H^2(x)]\upd y\,.
\end{align}
\end{subequations}
As $\rho^{+}(x,y)-\rho^{-}(x,y)=-2 \rho_0 \psi_0(x,y)$ , integrating Eq.\eqref{eq:QP} over 
$\int_0^L dx$, and imposing fluid incompressibility we get $\partial_{x}P=\frac{\bar{h}^3\Delta P}{(\eta\, L \;H^3(x) \,H_3)}$, where $H_3$ is defined in Eq.\eqref{eq:H3}.
This magnitude is a dimensionless geometrical measure for the corrugation of the {\it lubricated deformed} microchannel. We find $H_{3}\ge 1$, with the equality holding when the channel is flat $[h(x)=\bar{h}]$.
Therefore $\mathcal{J}_{q} (x)$ can be rewritten as
\begin{align}
\mathcal{J}_{q} (x) \equiv& - \gamma_0 \beta ze \frac{\Delta P}{2\eta L k_{0}^{2}}\frac{\bar{h}^3}{H_{3} H^2(x)}\times\nonumber\\
&\left[ \int_{-h(x)}^{h(x)} dy \psi_0(x,y)[y^2-H^2(x)]\right]\,.
\end{align}
By integrating, the last expression can be rewritten as 
\begin{equation}\label{eq:jq}
\mathcal{J}_{q} (x) \equiv - \gamma_0 \beta ze \frac{\Delta P}{2\eta L k_{0}^{2}}\frac{\bar{h}^3 \left[4 h(x) \psi_0(x,h(x)) -2 \overline{\psi_0}(x) \right]}{H_{3}\, H^3(x)}
\end{equation}
We now proceed as follows: from \eqr{eq:lmbd-eps-int} we derive expressions for $ \gamma_{f}(x)$ and $ \xi_{f}(x)$
in terms of the fluxes $J_{c}$, $J_{q}$, and $Q$ [cf. Eqs.~\eqref{eq:cf} and \eqref{eq:qf}]:
\begin{align}
   &\gamma_{f}(x)=\gamma_{f}(0)-\frac{J_{c}'}{2D}\int_0^x\!\frac{\upd x'}{h(x')}
		 -\frac{J_{q}'}{4D}\int_0^x\!\frac{\beta ze\overline{\psi_0}(x')}{h^{2}(x')}\,\upd x'\,,\label{eq:cf}
 \end{align}
 \begin{align}
   \xi_{f}(x)&=\xi_{f}(0)-\frac{J_{q}}{2D}\int_0^x\!\frac{\upd x'}{h(x')}- \frac{J_{c}'}{4D}\int_0^x\!\frac{\beta ze\overline{\psi_0}(x')}{h^{2}(x')}\,\upd x'\nn
	      &\quad+\frac{1}{2D}\int_0^x\!\frac{\mathcal{J}_{q} (x')}{h(x')}\,\upd x' \,.\label{eq:qf}
\end{align}
Using Eq.\ref{eq:jq} we calculate the last term in Eq.\ref{eq:qf} 
\begin{align}
\int_0^L\frac{\mathcal{J}_{q} (x)}{h(x)}\,\upd x
&=\gamma_0\frac{2 \Delta P}{ k_{0}^2 \eta L}\frac{\bar{h}^3}{H_{3}}\beta ze\times\\
&\int_0^L \upd x
\frac{4 h(x) \psi_0(x,h(x)) - 2 \overline{\psi_0}(x)}{h(x)H^3(x)} \nonumber\,,
\end{align}
Evaluating \eqr{eq:qf} at $x=L$, a term containing $\int_{0}^{L}\upd x\,\mathcal{J}(x')/h(x')$ appears. With the above two equations
we find 
\begin{align}\label{eq:app_theta3}
\int_0^L\frac{\mathcal{J}_{q} (x)}{h(x)}\,\upd x
&=2\gamma_0\frac{\Delta P}{k_{0}^2 \eta }\Phi \Upsilon_{3}\,,  
\end{align}
where we define
\begin{subequations}\label{eq:ips-def}
\begin{align}
H_{1}&\equiv \frac{\bar{h}}{L}\int_{0}^{L}\frac{1}{h(x)}\,\upd x\label{eq:H1}\,,\\
 \Phi &\equiv\beta ze\times\begin{cases}
                         \zeta& \quad\quad\text{conductive walls}\,,\\
                         \displaystyle{\frac{\sigma}{\epsilon k_{0}}}& \quad\quad\text{dielectric walls}\,,
   \end{cases}\\                        
\Upsilon_{1}&\equiv \frac{\bar{h}}{ H_{1}L}\int_0^L \!\!\upd x\,\frac{\beta ze\overline\psi_0(x)}{2 h^{2}(x)\Phi}\,,\label{eq:Upsilon1}\\
\Upsilon_{3}
&\equiv \frac{\bar{h}^3}{H_{3}}\frac{1}{L}\int_0^L \upd x
\frac{4 h(x) \psi_0(x,h(x)) - 2 \overline{\psi_0}(x)}{h(x)H^3(x)} \,
\,.\label{eq:Upsilon3}
\end{align}
\end{subequations}
Similar to $H_{3}$, $H_{1}$ is a geometrical measure of the channel.
$\Phi/(\beta ze)$ is a magnitude that equals the surface potential $\psi^{\zeta}(x,h(x))$ for conducting walls, while for dielectric surfaces it differs from $\psi^{\sigma}(x,h(x))$ by a factor  $\coth[k_{0}h(x)]$.
And finally, the $\Upsilon$ functions are related to the elements of the Onsager matrix corresponding to the out-of-equilibrium transport along the corrugated lubricated microchannel:
\begin{align}\label{eq:onsagermatrix}
 \left( \begin{array}{c} \displaystyle{J_{q}} \\ J'_{c} \\ Q  \end{array} \right) 
 =& \begin{pmatrix} \mathcal{L}_{11} & \mathcal{L}_{12} &\mathcal{L}_{13}\\ 
			\mathcal{L}_{21} & \mathcal{L}_{22} & 0 \\ 
			\mathcal{L}_{31}& 0 & \mathcal{L}_{33}\end{pmatrix} 
 \left( \begin{array}{c} ze\displaystyle{\Delta V}\\\displaystyle{\Delta \bar\mu}\\ \displaystyle{\Delta P} \end{array} \right)\frac{1}{L}\,,
\end{align}
 where the coefficients read
\begin{subequations}\label{eq:onsager_coeff}
 \begin{align}
\mathcal{L}_{11}&=\mathcal{L}_{22}=-2\gamma_0\frac{\bar h}{H_1}\frac{\mu_\text{ion}}{ze}\,,\label{eq:L11=L22}\\
\mathcal{L}_{12}&=\mathcal{L}_{21}=- 2\gamma_0\frac{\bar h}{H_1}\frac{\mu_\text{ion}}{ze}\Phi\Upsilon_{1}\,, \\
\mathcal{L}_{13}&=\mathcal{L}_{31}= - 2\gamma_0\frac{\bar h}{H_1}\frac{1}{\eta k_0^2}\Phi\Upsilon_{3}\,,\\
&\hspace{0.5cm}\mathcal{L}_{33}=-\frac{2}{3H_{3}}\frac{\bar{h}^3}{\eta }\,.
\end{align}
\end{subequations}

The Onsager matrix given in Eq.\ref{eq:onsagermatrix} is symmetric and it relates the charge flow $\mathcal{J}_{q} (x)$, the solute flow $\mathcal{J}^{'} (x)$ and the volumetric fluid flow $Q$,  to the three thermodynamic forces
$\frac{ze\displaystyle{\Delta V}}{L}$, $\frac{\displaystyle{\Delta \bar\mu}}{L}$, $\frac{\displaystyle{\Delta P}}{L}$ through the four independent nonzero transport coefficients $\mathcal{L}_{11}$,  $\mathcal{L}_{33}$, $\mathcal{L}_{12}$ and $\mathcal{L}_{13}$.
The last two are non zero only when the channel walls are charged ($\Phi \neq 0$).  On the other hand, the off-diagonal elements $\mathcal{L}_{23}$ (and $\mathcal{L}_{32}$) are zero for two reasons: on the one hand the linearization of the Poisson-Boltzmann equation (Debye-H\"uckel approximation). On the other hand in the current model we have disregarded additional interactions between ions and walls (such as van der Walls interactions) which may lead to diffusioosmotic fluxes.  Within this approach, the diagonal terms $\mathcal{L}_{11},\mathcal{L}_{22}$ are controlled solely by $H_1$. Therefore they do not depend on $H(x)$ and hence on the slip length. This is peculiar of the Debye-H\"uckel regime within which terms proportional to $\psi^2$ are disregarded. In the (non-linear) Poisson-Boltzmann regime corrections to these terms shall be accounted for~\cite{Vinogradova2021}.

Concerning the other terms, we note that Eqs.\eqref{eq:H3} and  (\ref{eq:Upsilon3}) depend on $H(x)$. Consequently, only $\mathcal{L}_{13}$  and $\mathcal{L}_{33}$ are sensitive to the presence of a slip length.
$\Upsilon_{3}$ as given in Eq.\ref{eq:Upsilon3} can be rewritten as follows
\begin{align}
\Upsilon_{3}
&=\frac{\bar{h}^3}{H^o_{3}}\frac{1}{L}\int_0^L 
\frac{4 h(x) \psi_0(x,h(x)) - 2 \overline{\psi_0}(x)}{h^4(x)}Z(x)\upd x\,,
\label{eq:Y_30-zeta}
\end{align}
with
\begin{align}
    H^o_3 &= \frac{\bar{h}^{3}}{L}\int_{0}^{L}\!\!\frac{1}{2h^3(x)}\,\upd x\,,\\
    Z(x)&=\frac{h^3(x)}{H^3(x)}\frac{H^o_{3}}{H_{3}}\,.
    \label{eq:def-zeta}
\end{align}
Therefore, the dependence of $\Upsilon_3$ on the slip length is accounted for by the zeta potential
\begin{align}
    \zeta_{pot}= \frac{4 h(x) \psi_0(x,h(x)) - 2 \overline{\psi_0}(x)}{h^4(x)}Z(x)\,, \label{eq:def-zetapot}
\end{align}
which is indeed the magnitude that is typically experimentally accessible~\cite{Vinogradova2021,Vinogradova2022,Vinogradova2023}. Interestingly, Eq.~\eqref{eq:def-zeta} shows that the corrections to the local zeta potential due to the local slip length, contained in $Z(x)$, are multiplicative and they do not depend on the dielectric or conducting character of the channel walls.

Making a small slip length approach, that is for $\lambda(x)\ll \bar{h}$, an insight of the complex interplay between corrugation and lubrication can be achieved. Under this approximation $Z(x)$ can be approximated by
\begin{align}
    Z(x) &\simeq \left(1-3\frac{\lambda(x)}{\bar{h}}\right)\left(1+3\frac{1}{L}\int_0^L\frac{\lambda(x)}{\bar{h}}dx\right)\nonumber\\
    &\simeq \left(1-3\frac{\lambda(x)}{\bar{h}}\right)\left(1+3\frac{\bar{\lambda}}{\bar{h}}\right)\,,
\end{align}
where we introduced $\bar{\lambda}\equiv 1/L\int_0^L\lambda(x)dx$. 
Accordingly, we have 
\begin{align}
\Upsilon_{3}
&= \frac{\bar{h}^3}{H^o_{3}}\left(1+3\frac{\bar{\lambda}}{\bar{h}}\right)\frac{1}{L}\times\\
&\int_0^L 
\frac{4 h(x) \psi_0(x,h(x)) - 2 \overline{\psi_0}(x)}{h^4(x)}\left(1-3\frac{\lambda(x)}{\bar{h}}\right)\upd x\,.\nonumber
\end{align}
Interestingly, for constant-section channels the last expression reduces to 
\begin{align}
\Upsilon_{3}
&\simeq  
 4\frac{ \bar{h} \psi_0(\bar{h}) - \overline{\psi_0}}{\bar{h}}\left[1-9 \frac{\bar{\lambda}^2}{\bar{h}^2}\right]\,,
\label{eq:Y_30-zeta-h0}
\end{align}
which implies that for constant-section channels the corrections to $\Upsilon_3$ are quadratic in the average slip length  and they lead to an overall reduction of $\Upsilon_3$. 
In contrast, for varying-section channels the contribution of the local slip length is weighted by the local electrostatic potential and hence it leads to non vanishing corrections to $\Upsilon_3$ that are linear in $\lambda(x)$. Therefore, nanochannels with varying geometry are more sensitive to variations in their surface properties as compared to their constant-section counterparts. In order to identify the relevant dimensionless parameters it is insightful to rewrite Eq.~\eqref{eq:Y_30-zeta} as
\begin{align}\label{eq:Y_30-zeta-adim}
\Upsilon^{\sigma,\zeta}_{3}
&=\frac{4}{H^o_{3}}\frac{1}{L}\Phi\times\\
&\int_0^L 
\left(\frac{\bar{h}^3}{h^3(x)}\right) \frac{\chi^{\sigma,\zeta}(x)}{k_0 h(x)}\left[\frac{k_0 h(x)}{\tanh(k_0 h(x))}-1\right] Z(x)\upd x\,,\nonumber
\end{align}
with
\begin{align}
\chi^{\sigma,\zeta}(x)=
    \begin{cases}
        1 & \text{dielectric walls}\,,\\
        \tanh (k_0 h(x)) & \text{conductive walls}\,.
    \end{cases}
\label{eq:def-chi}
\end{align}
According to Eqs.~\eqref{eq:Y_30-zeta-adim},\eqref{eq:def-chi} we expect that in the "thin microchannel regime" i.e., for $k_0 \bar{h}<1$ corrections due to the slip length are more prominent for the dielectric case than for the conducting case since, for the latter, the function $\chi$ provides an enhanced decay with $k_0 \bar{h}$ as indeed it happens in the absence of slip  (see Ref.~\cite{PM19}).

\section{Results}\label{sec:Results}
\begin{figure}
    \centering
    \includegraphics[scale = 0.5]{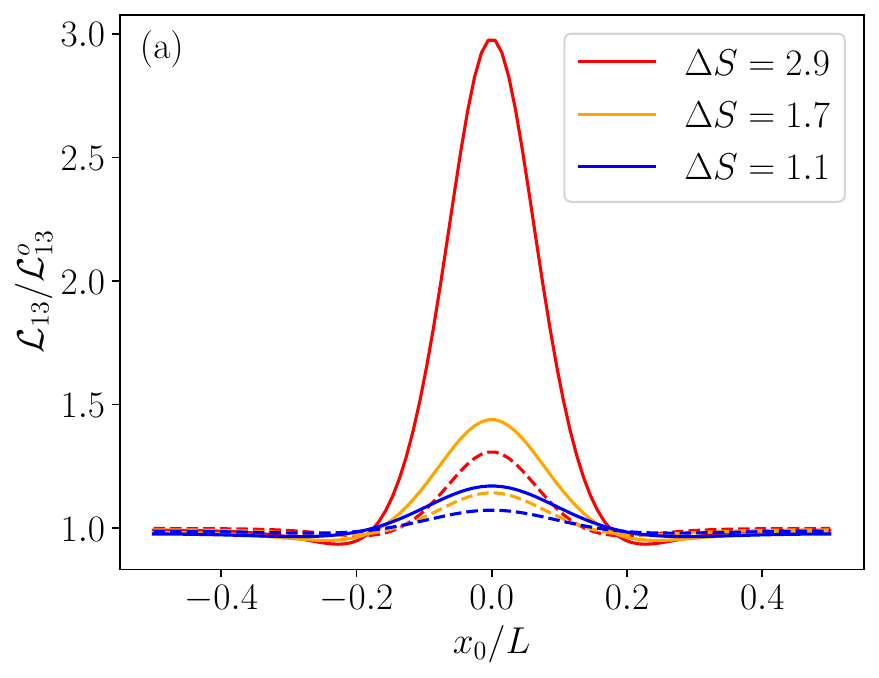}
    \includegraphics[scale = 0.5]{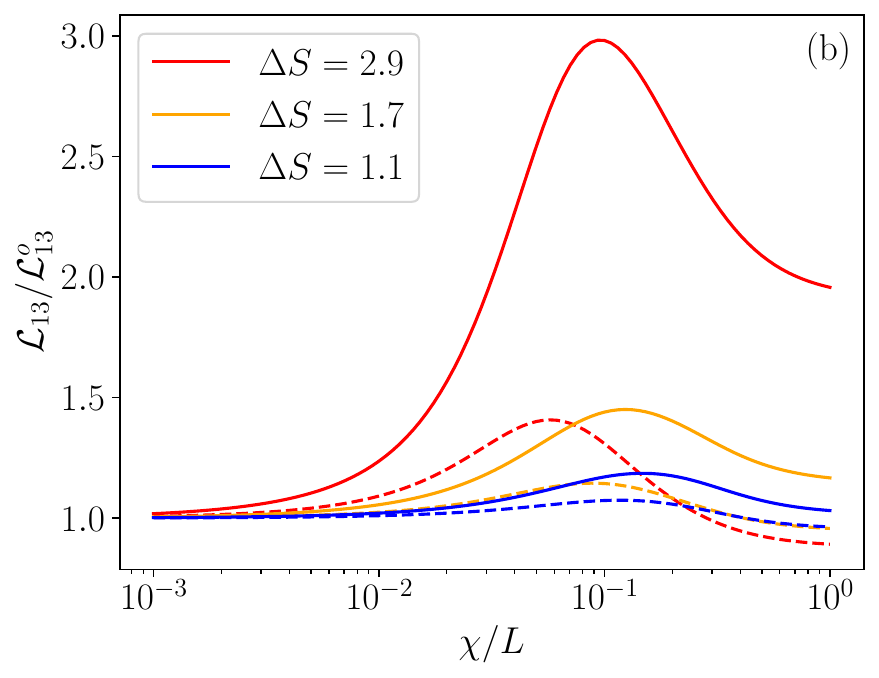}
    \caption{a) $\LL$, normalized by its value with $\ell_0=0$, as a function of the position, $x_0/L$, of its maximum expressed in units of the channel period, $L$ with $k_0 \bar{h}=1$, $\chi/L=1/10$. and $\Delta S$ as reported in the legend. Solid (dashed) lines stand for conducting (dielectric) walls. b) $\LL$, normalized by its value with $\ell_0=0$, as a function of the width of the slip length profile, $\chi/L$, in units of the period of the channel, $L$, with $k_0 \bar{h}=1$, $x_0/L=0$. and $\Delta S$ as reported in the legend. Solid (dashed) lines stand for conducting (dielectric) walls.}
    \label{fig:x0_chi}
\end{figure}

As we mentioned above, we focus on a nanochannel with a simple shape defined by Eq.~\eqref{eq:hx}. We consider a slip length with a Gaussian profile
\begin{align}\label{eq:lambda}
    \ell(x) = \ell_0 e^{-\frac{(x-x_0)^2}{\chi^2}}\,,
\end{align}
where ${\ell_0}$, $\chi$ and $x_0$ are the depth, width and position of the center, respectively (see Fig.\ref{sketch}). 
The dimensionless amplitude $h_1$ gives a sense of the nanochannel corrugation, however in the following to quantify this feature we prefer to use the corresponding “entropic barrier” defined as 
\begin{align}
\Delta S=\ln{\frac{1+h_1}{1-h_1}}\,,   
\end{align}
\begin{figure}[h]
    \centering
    \includegraphics[scale = 0.5]{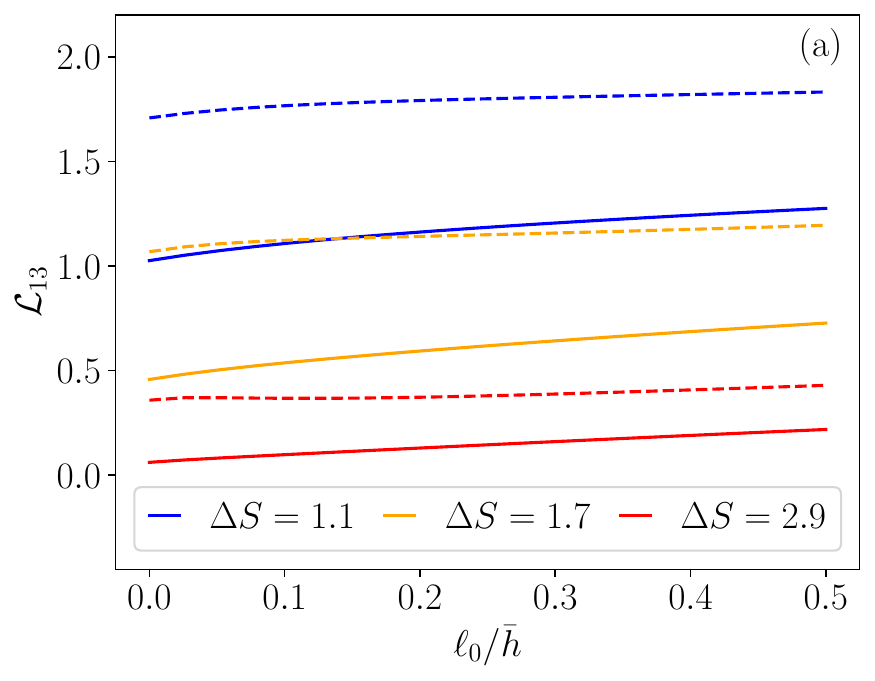}
    \includegraphics[scale = 0.5]{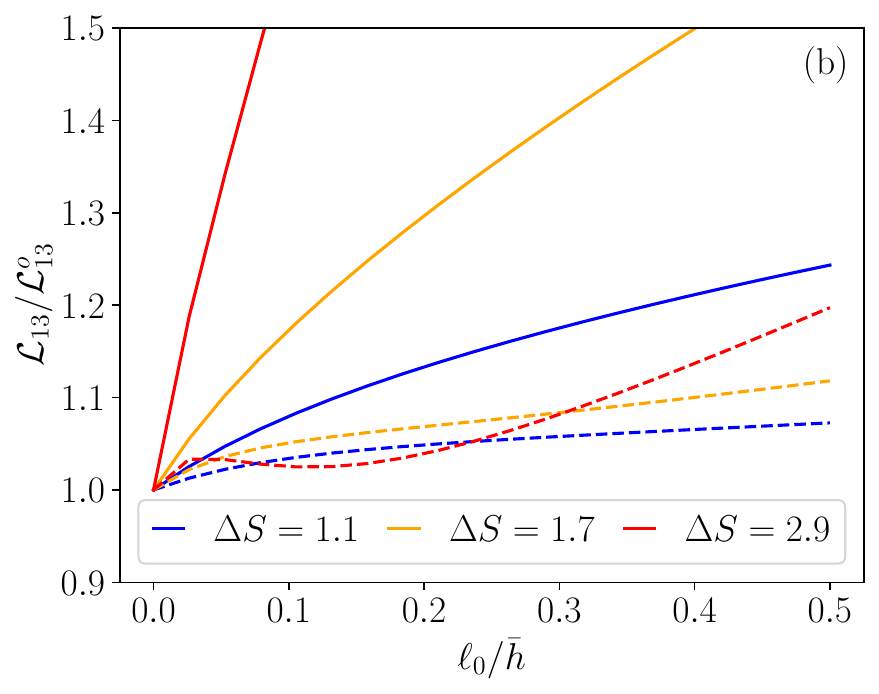}
    \caption{a) $\mathcal{L}_{13}$ as a function of the amplitude of the slip length, $\ell_0$, normalized by the average microchannel section, $\bar{h}$ for $k_0 \bar{h} = 1$ and weak confinement, $\Delta S= 1.1$ (equivalent to $h_1 = 0.5$), mild confinement, $\Delta S= 1.7$ (equivalent to $h_1 = 0.7$), and strong confinement, $\Delta S = 2.9$ (equivalent to $h_1 = 0.9$), as in the legend. The slip length profile, Eq.~\eqref{eq:lambda}, is characterized by $\chi/L= 0.2$ and $x_0=0$. Solid (dashed) lines stand for conducting (dielectric) walls. b) same data as in panel a) but normalized by the value obtained for $\ell_0 =0$.}
    \label{fig:L13_z}
\end{figure}
which captures the difference in entropy of an uncharged point particle confined within the channel~\cite{Zwanzig,entropic2,MPR2013}.
Fig.~\ref{fig:L13_z}a shows that, for both dielectric and conducting channel walls, $\mathcal{L}_{13}$ increases upon growing the magnitude of the slip length, $\ell_0$ for weak confinement, $\Delta S = 1.1$, mild confinement, $\Delta S = 1.7$, and strong confinement, $\Delta S = 2.9$.
In order to better assess the relative influence of the slip length on $\LL$ we focus on the ratios  \begin{align}
\frac{\mathcal{L}_{13}}{\mathcal{L}_{13}^o} = \frac{\Upsilon_{13}}{\Upsilon_{13}^o} = \frac{\int_0^L 
\frac{4 h(x) \psi_0(x,h(x)) - 2 \overline{\psi_0}(x)}{h(x)H^3(x) H_3} dx}{\int_0^L 
\frac{4 h(x) \psi_0(x,h(x)) - 2 \overline{\psi_0}(x)}{h^4(x) H_3^o} dx}\,,
\end{align}
where the upper index $o$ refers to the case of vanishing slip length. In particular, 
for the dielectric case we have
\begin{align}
4 h(x) \psi_0(x,h(x)) - 2 \overline{\psi_0}(x) = \frac{4}{k_0}\left[\frac{k_0 h(x)}{\text{tanh}(k_0 h(x))}-1\right]\,,
\label{eq:LL-diel}
\end{align}
while for the conducting case we have 
\begin{align}
4 h(x) \psi_0(x,h(x)) - 2 \overline{\psi_0}(x) = 4\frac{k_0 h(x)-\text{tanh}(k_0 h(x))}{k_0}\,.
\label{eq:LL-cond}
\end{align}
As shown in Fig.~\ref{fig:L13_z}b, conducting walls lead to a more sensitive dependence of $\LL$ on the slip length reaching an increase of almost $\sim 4$-fold, as compared to the case of dielectric walls which display up to $\sim 1.3$-fold increase in $\LL$. 
The effective slip length profile that we introduce in Eq.~\eqref{eq:lambda} also depends on the position of its maximum, $x_0$ and width, $\chi$. 
Indeed, Fig.\eqref{fig:x0_chi} reveals that the localization of the slip length is one of the important features to take into account to improve transport  assisted by hydrophobic walls. In fact, $\LL$ has a non-monotonic dependence on the position of the maximum of the slip length profile, $x_0$ with the presence of local maxima and minima located at almost the same positions for both conducting and dielectric channel walls. 
In particular, for both conducting and dielectric walls, $\LL$ attains a maximum for $x_0/L=0$, i.e. when the slip length is maximum at the microchannel bottleneck. Upon shifting the slip length profile away from the bottleneck, $\LL$ reduces until it reaches a minimum for which $\LL/\LLo<1$ i.e., the slip length hinders the transport. Upon moving the slip length profile further away $\LL/\LLo\rightarrow 1$ and the slip length almost plays no role on $\LL$.  

\begin{figure}
\includegraphics[scale = 0.5]{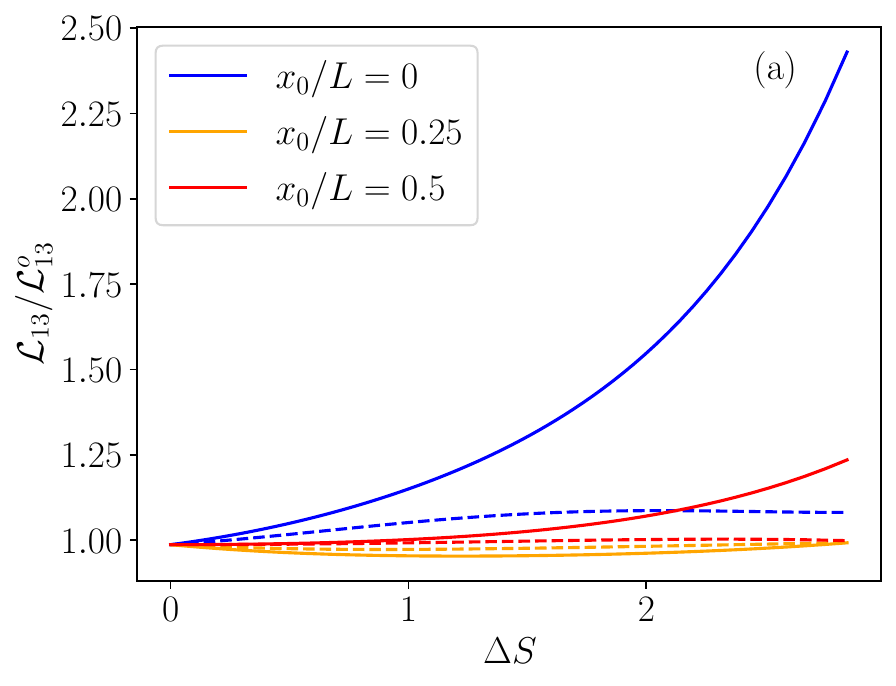}
\includegraphics[scale = 0.5]{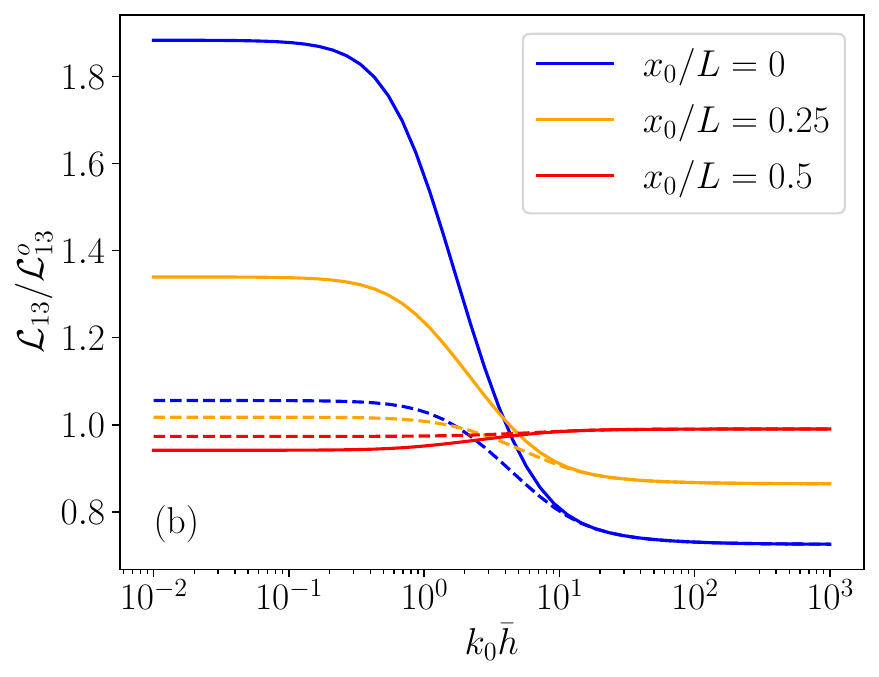}
\caption{$\mathcal{L}_{13}/\mathcal{L}^{o}_{13}$ versus $\Delta S$ ($ k_0\bar{h}=1$) (a) and $k_0\bar{h}$ ($\Delta S=4.35$) (b). Different curves correspond to $x_0/L = 0, 0.1, 0.5$ (blue, orange, and red respectively). In all cases full/dashed lines corresponds to conductive/dielectric cases. Other parameters are $\chi /L=0.2$ and $\ell_0/\bar{h}=0.3$.\label{fig:L13NhoDs}} 
\end{figure}

So far we have discussed the dependence of $\LL$ on the properties of the slip length profile, such as its magnitude ($\ell_0$), location ($x_0$) and width ($\chi$). However, as already anticipated in the previous figures, $\LL/\LLo$  depends also on the channel geometry, captured by $\Delta S$. 
Fig.~\ref{fig:L13NhoDs}a shows the ratio $\LL/\LLo$ as a function of $\Delta S$ for the case of a slip length centered at three different positions: the bottleneck ($x_0=0$), an intermediate position ($x_0=0.2$) and the widest section of the channel ($x_0=0.5$).
In Fig.~\ref{fig:L13NhoDs}a we can identify three main scenarios. First, for all values of $\Delta S$ the major deviation of $\LL/\LLo$ from unity occurs for a slip length profile centered at the bottleneck, $x_0 =0$. In this case $\LL/\LLo>1$ as $\Delta S$ increases. 
Second, for conducting channel walls and for $x_0=0.5$, $\LL/\LLo$ displays a non monotonic behavior, becoming even smaller than one. 
In contrast, for dielectric channel walls, when the slip length profile is centered away from the microchannel bottleneck, $x_0/L \neq 0$,  $\LL/\LLo$ is almost independent of $\Delta S$. 

Finally, we analyze the dependence of $\LL$ on the inverse Debye length, $k_0$.
Fig.~\ref{fig:L13NhoDs}b shows the dependence of $\LL/\LLo$ on  $k_0 \bar{h}$. In this case we note a qualitative difference depending on the location of the maximum of the slip length, $x_0$. In particular, for $x_0/L=0$ the curves display a monotonic decay of $\LL$ upon increasing $k_0 \bar{h}$ for both dielectric and conducting channel walls. While for $k_0 \bar{h} \ll 1 $ we observe an enhanced sensitivity of $\LL/\LLo$ for conducting channels as compared to dielectric channels, for $k_0 \bar{h}\gg 1$ the data for the two channel walls converge onto a single curve. 
This behavior for large values of $k_0 \bar{h}$ can be predicted by the functional form of Eqs.~\eqref{eq:LL-diel},\eqref{eq:LL-cond} which already show that the two integrands differ by a numerical prefactor that is $\text{tanh}(k_0 \bar{h})$ and hence they approach each other for $k_0 
\bar{h}\gg 1$.
It is interesting to note that for $x_0/L=0.5$ and for $k_0 \bar{h} \leq 1$,  $\LL/\LLo<1$. This is quite counter-intuitive since one would expect an enhancement of the transport upon increasing the slip. However, the results shown in Fig.~\ref{fig:L13NhoDs}b imply that in channels with an average height smaller than the Debye length, $k^{-1}_0$, adding a slip length profile worsens the transport performance as compared to the bare channel.

So far we discussed the influence of the slip length profile on the off-diagonal Onsager coefficient $\LL$. Now we move our attention to the only other Onsager coefficient that is sensitive to the local slip, namely $\TT$.
\begin{figure}
    \centering
    \includegraphics[scale = 0.5]{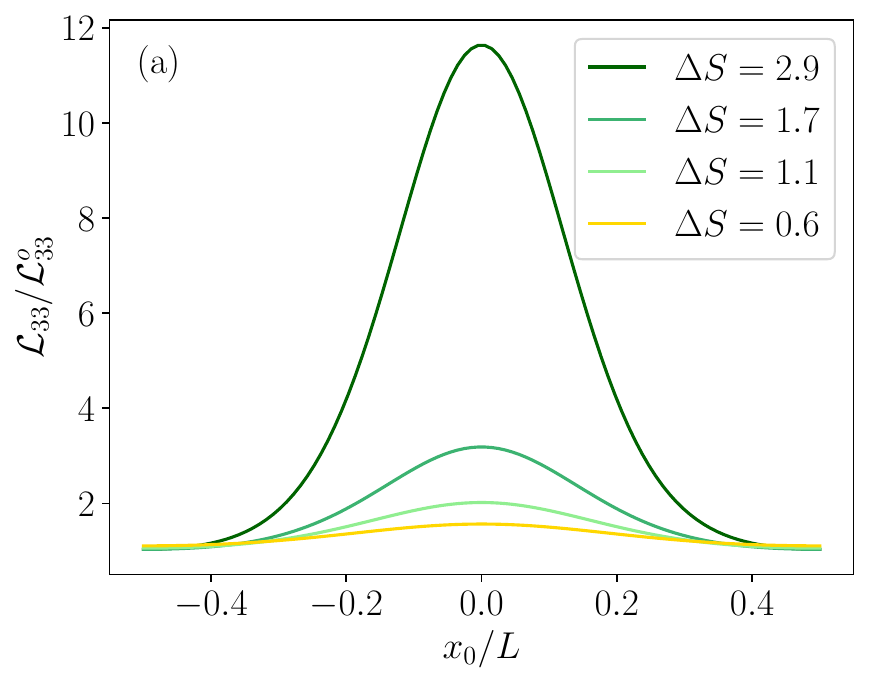}
    \includegraphics[scale = 0.5]{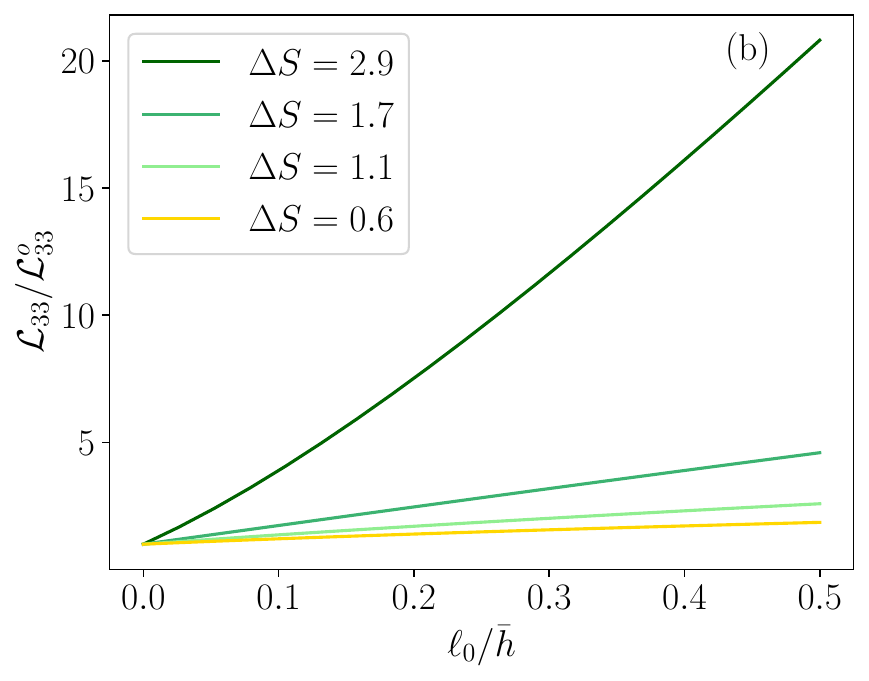}
    \includegraphics[scale = 0.5]{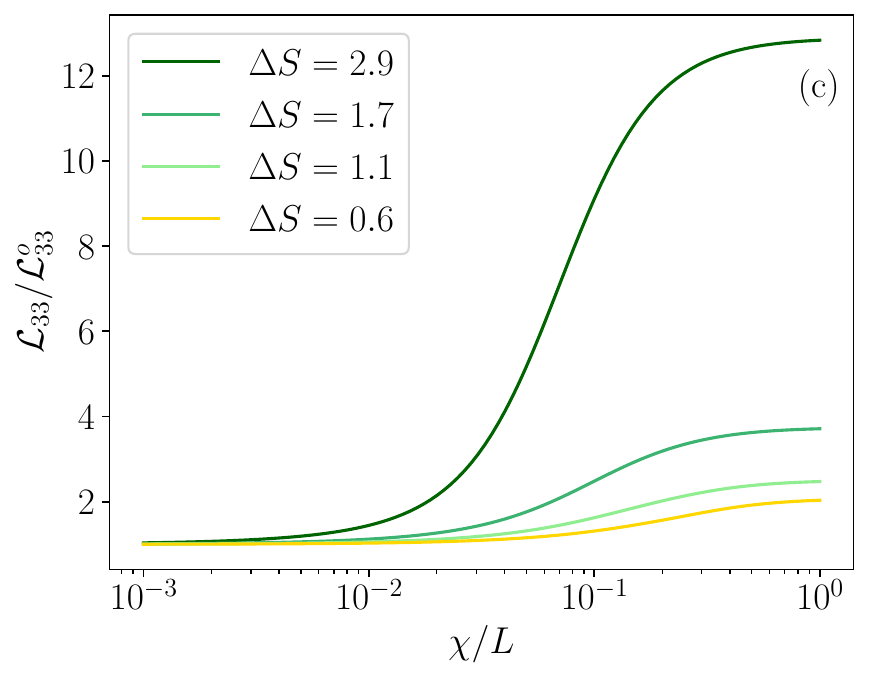}
    \caption{(a) $\mathcal{L}_{33}/\mathcal{L}^o_{33}$ versus $x_0/L$ with $\chi/L=0.2$ and $\ell_0/\bar{h}=0.3$, (b) $\mathcal{L}_{33}/\mathcal{L}^o_{33}$ versus $\ell_0/\bar{h}$ with $\chi/L=0.2$ and $x_0=0$. (c) $\mathcal{L}_{33}/\mathcal{L}^o_{33}$ versus $\chi/L$ with $\ell_0/\bar{h}=0.3$ and $x_0=0$.}
    \label{fig:LL33}
\end{figure}
Fig.~\ref{fig:LL33}a shows the dependence of $\TT/\TTo$ on the position of the maximum of the slip length profile. Interestingly, $\TT/\TTo$ displays a maximum for $x_0/L=0$, as we discussed also for $\LL/\LLo$. However, at variance with $\LL/\LLo$,  $\TT/\TTo$ does not have minima close to the maximum and its value decreases monotonically till its minimum value attained for $x_0/L=0.5$ i.e., when the maximum of the slip length profile is located at the widest section of the channel. 
As expected, $\TT/\TTo$ grows monotonically upon increasing both the amplitude of the slip length profile, $\ell_0$ as well as the width of the slip length profile, $\chi$.

Eqs.\eqref{eq:onsagermatrix} show that cross phenomena may affect the fluxes, e.g.
a solvent or solute current can be prompted by applying an external field on the electrolytes or a pressure difference along the channel. Although the linear set of equations of Eqs.\eqref{eq:onsagermatrix} leads to non-trivial couplings between currents, it is interesting to analyze some particular cases of interest in micro and nanofluidic applications.

As we discussed above, there are different possibilities to enhance $ \frac{\TT}{\TTo}$ and $ \frac{\mathcal{L}_{13}}{\mathcal{L}^o_{13}}$, and so the currents, that can be used to take advantage of the existence of slip boundary layers found in many applications. For example, one of the challenges in micro/nanofluidic-based thermal devices is to remove as much heat as possible by forced convection of the fluid, and at the same time to minimize power used to pump the fluid.
Our results show that considering a pressure-driven flow control mediated only by pumps, that is $\Delta P \neq 0$ , $\Delta \mu =0 $ and $\Delta V =0$, the fluxes are given by
\begin{align}
\begin{array}{c} J_{q}= \mathcal{L}_{13} \Delta P=  \frac{\mathcal{L}_{13}} {\mathcal{L}^o_{13}}  J_{q}^o\,,
\\
\\Q=  \mathcal{L}_{33}\Delta P =  \frac{\mathcal{L}_{33}}{\mathcal{L}^o_{33}}  Q^o\,.
\end{array} 
\end{align}
where $J_q^o$ and $Q^o$ are the electric current and the volumetric fluid flow in the same nanoochannel with $\ell_0=0$. 
Therefore, according to our results, a strategy to reduce the power consumed by pumping is to customize the design of the micro/nanochannels, making use of the joint effect of geometry and hydrophobic coatings.
Similarly, in the case of electro-osmotic flow, 
$\Delta V \neq 0$  and  $\Delta \mu =\Delta P=0$ we have 
\begin{align}
Q=  \mathcal{L}_{13}\Delta V =  \frac{\mathcal{L}_{13}}{\mathcal{L}^o_{13}}  Q^o\,.    
\end{align}

\section{Conclusions}\label{sec:Conclusions}

We studied the transport of a z-z electrolyte embedded in a varying smooth-section nanochannel with inhomogeneous slip. 
To keep the model simple, we have restricted ourselves to a linear response and to the Debye-H\"uckel regime within which the Onsager matrix can be derived. 
We also assumed a length scale separation between the longitudinal and transverse length scales so that we can exploit the lubrication approximation and reduce the model to an effective $1D$ system.
Within such a regime two off-diagonal elements, $\mathcal{L}_{23}=\mathcal{L}_{32}$, are zero. Of the remaining coefficients, only two are sensitive to the  local slip: the off-diagonal coefficient controlling the onset of electric currents when a pressure drop is applied (or alternatively the onset of fluid flow by applying an electrostatic potential drop) $\LL=\mathcal{L}_{31}$ and the diagonal coefficient $\TT$ controlling the magnitude of fluid flow when a pressure drop is applied. 
To quantify these effects, we focused on the case of a simple sinusoidal nanochannel with a Gaussian slip length profile. 
For what concerns $\LL$, our results show a non-trivial dependence on both the position of the maximum of the slip length profile, $x_0/L$, as well as on the width of the slip length profile, $\chi/L$. Interestingly, our results show that $\LL$ is maximized when the maximum of the slip length profile is located at the channel bottleneck and we also find local minima for $x_0/L\simeq \pm 0.2$. Surprisingly, we also find that the dependence of $\LL$ on $\chi$ is non-monotonic. This one the one hand is counterintuitive since one might expect a monotonic increase in $\LL$ as the slip region is enlarged. On the other hand, the non-monotonic dependence of $\LL$ on $\chi/L$ identifies an optimal value of $\chi/L\simeq 0.1$ for which $\LL$ is maximized, see Fig.\ref{fig:x0_chi}. 
So far we discussed the ratio $\mathcal{L}_{13}/\mathcal{L}_{13}^o$. 

Accordingly, since $\mathcal{L}_{13}^o$ displays a non-trivial dependence on $\Delta S$~\cite{PM19} it is interesting to discuss the magnitude of $\mathcal{L}_{13}$ per se. In the absence of slip, it was shown in Ref.~\cite{PM19} that $\mathcal{L}^o_{13}$ decreases with $\Delta S$, so the electroosmotic and electric currents induced by a $\Delta V$ and a $\Delta P$, respectively, decrease with $\Delta S$. 
However, as shown in Fig.~\ref{fig:L13_z}, $\LL$ increases as the slip length $\ell_0$ increases.
Therefore, the increase in $\LL$ due to the slip length contrasts with the decrease in $\LL$ due to the geometry, which is captured by $\Delta S$. So it looks as if the presence of a finite and localized slip length reduces the effect of the geometry (which actually reduces the magnitude of $\LL$).  In contrast, the behavior of $\TT$ is much more intuitive as it shows monotonic trends with increasing size and width of the slip profile and, as expected, a maximum for $x_0/L=0$, i.e. when the slip region is centered around the nanochannel bottleneck.

\section{Acknowledgment}
MFC thanks PIP-CONICET 11220200101599CO and CyTUNGS 30/1161. JH and PM acknowledge funding by the Deutsche Forschungsgemeinschaft (DFG, German Research Foundation) Project-ID 416229255-SFB 1411.

\appendix
\section{Equivalence between the Navier slip condition and Eq.~\eqref{eq:def-lambda}}\label{app:A}
Within the lubrication approximation
\begin{align}
    h(x) \ll L\,,
\end{align}
the Navier slip condition reads
\begin{align}
    v|_{y=\pm h(x)} = \ell(x) \partial_y v|_{y=\pm h(x)}\,,
\end{align}
where $\ell(x)$ is the slip length. 
In a straight channel the Navier boundary conditions lead to the velocity profile
\begin{align}
    v_\ell(x,y) = -\frac{\nabla P(x)}{\eta}\left(y^2-h^2(x)-2\ell(x) h(x)\right)\,.
\end{align}
In contrast, the no-slip boundary condition at the fictitious boundary $H(x)=h(x)+\lambda(x)$ (Eq.~\eqref{eq:def-lambda}) leads to 
\begin{align}
    v_\ell(x,y) = -\frac{\nabla P(x)}{\eta}\left(y^2-(h(x) \pa{+}\lambda(x))^2\right)\,,
\end{align}
which is the standard Poiseuille flow between parallel walls located at $H(x)=h(x)+\lambda(x)$.
By equating the last two expressions we get the relation between $\lambda$ and $\ell$:
\begin{align}
    \lambda(x) = h(x)\left(\sqrt{1+2\frac{\ell(x)}{h(x)}}-1\right)\,,
    \label{eq:lambda-map}
\end{align}
which shows that the solution associated to the fictitious boundary, $H(x) = h(x) + \lambda(x)$, can always be mapped to the solution of the usual Navier slip length, $\ell(x)$, via Eq.~\eqref{eq:lambda-map}. Moreover, for $\ell(x) \ll h(x)$ we have
\begin{align}
    \lambda(x) \simeq \ell(x)\,.
\end{align}
i.e., for $\ell \ll h$ the distance between the fictitious boundary and the microchannel walls $\lambda$  approaches the Navier slip $\ell$.

\bibliography{referencias.bib}

\begin{thebibliography}{61}%
\makeatletter
\providecommand \@ifxundefined [1]{%
 \@ifx{#1\undefined}
}%
\providecommand \@ifnum [1]{%
 \ifnum #1\expandafter \@firstoftwo
 \else \expandafter \@secondoftwo
 \fi
}%
\providecommand \@ifx [1]{%
 \ifx #1\expandafter \@firstoftwo
 \else \expandafter \@secondoftwo
 \fi
}%
\providecommand \natexlab [1]{#1}%
\providecommand \enquote  [1]{``#1''}%
\providecommand \bibnamefont  [1]{#1}%
\providecommand \bibfnamefont [1]{#1}%
\providecommand \citenamefont [1]{#1}%
\providecommand \href@noop [0]{\@secondoftwo}%
\providecommand \href [0]{\begingroup \@sanitize@url \@href}%
\providecommand \@href[1]{\@@startlink{#1}\@@href}%
\providecommand \@@href[1]{\endgroup#1\@@endlink}%
\providecommand \@sanitize@url [0]{\catcode `\\12\catcode `\$12\catcode
  `\&12\catcode `\#12\catcode `\^12\catcode `\_12\catcode `\%12\relax}%
\providecommand \@@startlink[1]{}%
\providecommand \@@endlink[0]{}%
\providecommand \url  [0]{\begingroup\@sanitize@url \@url }%
\providecommand \@url [1]{\endgroup\@href {#1}{\urlprefix }}%
\providecommand \urlprefix  [0]{URL }%
\providecommand \Eprint [0]{\href }%
\providecommand \doibase [0]{http://dx.doi.org/}%
\providecommand \selectlanguage [0]{\@gobble}%
\providecommand \bibinfo  [0]{\@secondoftwo}%
\providecommand \bibfield  [0]{\@secondoftwo}%
\providecommand \translation [1]{[#1]}%
\providecommand \BibitemOpen [0]{}%
\providecommand \bibitemStop [0]{}%
\providecommand \bibitemNoStop [0]{.\EOS\space}%
\providecommand \EOS [0]{\spacefactor3000\relax}%
\providecommand \BibitemShut  [1]{\csname bibitem#1\endcsname}%
\let\auto@bib@innerbib\@empty
\bibitem [{\citenamefont {Ghaderinezhad}\ \emph {et~al.}(2020)\citenamefont
  {Ghaderinezhad}, \citenamefont {Koydemir}, \citenamefont {Derek~Tseng},
  \citenamefont {Liang}, \citenamefont {Ozcan},\ and\ \citenamefont
  {Tasoglu}}]{Ghader2020}%
  \BibitemOpen
  \bibfield  {author} {\bibinfo {author} {\bibfnamefont {F.}~\bibnamefont
  {Ghaderinezhad}}, \bibinfo {author} {\bibfnamefont {H.~C.}\ \bibnamefont
  {Koydemir}}, \bibinfo {author} {\bibfnamefont {D.~K.}\ \bibnamefont
  {Derek~Tseng}}, \bibinfo {author} {\bibfnamefont {K.}~\bibnamefont {Liang}},
  \bibinfo {author} {\bibfnamefont {A.}~\bibnamefont {Ozcan}}, \ and\ \bibinfo
  {author} {\bibfnamefont {S.}~\bibnamefont {Tasoglu}},\ }\bibfield  {title}
  {\enquote {\bibinfo {title} {Sensing of electrolytes in urine using a
  miniaturized paper-based device},}\ }\href {\doibase
  10.1038/s41598-020-70456-6} {\bibfield  {journal} {\bibinfo  {journal}
  {Scientific Reports}\ }\textbf {\bibinfo {volume} {10}} (\bibinfo {year}
  {2020}),\ 10.1038/s41598-020-70456-6}\BibitemShut {NoStop}%
\bibitem [{\citenamefont {Lapizco-Encinas}(2021)}]{Lapizco2021}%
  \BibitemOpen
  \bibfield  {author} {\bibinfo {author} {\bibfnamefont {B.~H.}\ \bibnamefont
  {Lapizco-Encinas}},\ }\bibfield  {title} {\enquote {\bibinfo {title}
  {Microscale nonlinear electrokinetics for the analysis of cellular materials
  in clinical applications: a review},}\ }\href {\doibase
  10.1007/s00604-021-04748-7} {\bibfield  {journal} {\bibinfo  {journal}
  {Microchimica Acta}\ }\textbf {\bibinfo {volume} {188}} (\bibinfo {year}
  {2021}),\ 10.1007/s00604-021-04748-7}\BibitemShut {NoStop}%
\bibitem [{\citenamefont {Reddy}, \citenamefont {Basha},\ and\ \citenamefont
  {Duraisamy}(2022)}]{Reddy2022}%
  \BibitemOpen
  \bibfield  {author} {\bibinfo {author} {\bibfnamefont {S.}~\bibnamefont
  {Reddy}}, \bibinfo {author} {\bibfnamefont {H.}~\bibnamefont {Basha}}, \ and\
  \bibinfo {author} {\bibfnamefont {P.}~\bibnamefont {Duraisamy}},\ }\bibfield
  {title} {\enquote {\bibinfo {title} {Entropy generation for peristaltic flow
  of gold-blood nanofluid driven by electrokinetic force in a microchannel.}}\
  }\href {\doibase 10.1140/epjs/s11734-021-00379-4} {\bibfield  {journal}
  {\bibinfo  {journal} {Eur. Phys. J. Spec. Top.}\ }\textbf {\bibinfo {volume}
  {231}},\ \bibinfo {pages} {2409} (\bibinfo {year} {2022})}\BibitemShut
  {NoStop}%
\bibitem [{\citenamefont {Gross}\ and\ \citenamefont
  {Osterle}(1968)}]{Gross1968}%
  \BibitemOpen
  \bibfield  {author} {\bibinfo {author} {\bibfnamefont {R.~J.}\ \bibnamefont
  {Gross}}\ and\ \bibinfo {author} {\bibfnamefont {J.~F.}\ \bibnamefont
  {Osterle}},\ }\bibfield  {title} {\enquote {\bibinfo {title} {Membrane
  transport characteristics of ultrafine capillaries},}\ }\href {\doibase
  10.1063/1.1669814} {\bibfield  {journal} {\bibinfo  {journal} {The Journal of
  Chemical Physics}\ }\textbf {\bibinfo {volume} {49}},\ \bibinfo {pages}
  {228--234} (\bibinfo {year} {1968})},\ \Eprint
  {http://arxiv.org/abs/https://pubs.aip.org/aip/jcp/article-pdf/49/1/228/18856765/228\_1\_online.pdf}
  {https://pubs.aip.org/aip/jcp/article-pdf/49/1/228/18856765/228\_1\_online.pdf}
  \BibitemShut {NoStop}%
\bibitem [{\citenamefont {Cheng}(2018)}]{chen2018}%
  \BibitemOpen
  \bibfield  {author} {\bibinfo {author} {\bibfnamefont {L.-J.}\ \bibnamefont
  {Cheng}},\ }\bibfield  {title} {\enquote {\bibinfo {title} {{Electrokinetic
  ion transport in nanofluidics and membranes with applications in bioanalysis
  and beyond}},}\ }\href {\doibase https://doi.org/10.1063/1.5022789}
  {\bibfield  {journal} {\bibinfo  {journal} {Biomicrofluidics}\ }\textbf
  {\bibinfo {volume} {12}},\ \bibinfo {pages} {021502} (\bibinfo {year}
  {2018})}\BibitemShut {NoStop}%
\bibitem [{\citenamefont {Zhou}\ and\ \citenamefont {Lei}(2020)}]{Zhou2020}%
  \BibitemOpen
  \bibfield  {author} {\bibinfo {author} {\bibfnamefont {Y.}~\bibnamefont
  {Zhou}}\ and\ \bibinfo {author} {\bibfnamefont {J.}~\bibnamefont {Lei}},\
  }\bibfield  {title} {\enquote {\bibinfo {title} {Bioinspired nanoporous
  membrane for salinity gradient energy harvesting},}\ }\href {\doibase
  10.1016/j.joule.2020.09.009} {\bibfield  {journal} {\bibinfo  {journal}
  {Joule}\ }\textbf {\bibinfo {volume} {4}},\ \bibinfo {pages} {2244} (\bibinfo
  {year} {2020})}\BibitemShut {NoStop}%
\bibitem [{\citenamefont {Sun}\ \emph {et~al.}(2023)\citenamefont {Sun},
  \citenamefont {Jiang}, \citenamefont {Hu}, \citenamefont {Song},\ and\
  \citenamefont {Li}}]{Sun2023}%
  \BibitemOpen
  \bibfield  {author} {\bibinfo {author} {\bibfnamefont {Y.}~\bibnamefont
  {Sun}}, \bibinfo {author} {\bibfnamefont {R.}~\bibnamefont {Jiang}}, \bibinfo
  {author} {\bibfnamefont {L.}~\bibnamefont {Hu}}, \bibinfo {author}
  {\bibfnamefont {Y.}~\bibnamefont {Song}}, \ and\ \bibinfo {author}
  {\bibfnamefont {M.}~\bibnamefont {Li}},\ }\bibfield  {title} {\enquote
  {\bibinfo {title} {Electrokinetic transport phenomena in nanofluidics and
  their applications},}\ }\href {\doibase
  https://doi.org/10.1002/elps.202300115} {\bibfield  {journal} {\bibinfo
  {journal} {Electrophoresis}\ }\textbf {\bibinfo {volume} {44}},\ \bibinfo
  {pages} {1756} (\bibinfo {year} {2023})}\BibitemShut {NoStop}%
\bibitem [{\citenamefont {Aluru}\ \emph {et~al.}(2023)\citenamefont {Aluru},
  \citenamefont {Aydin}, \citenamefont {Bazant}, \citenamefont {Blankschtein},
  \citenamefont {Brozena}, \citenamefont {de~Souza}, \citenamefont {Elimelech},
  \citenamefont {Faucher}, \citenamefont {Fourkas}, \citenamefont {Koman},
  \citenamefont {Kuehne}, \citenamefont {Kulik}, \citenamefont {Li},
  \citenamefont {Li}, \citenamefont {Li}, \citenamefont {Majumdar},
  \citenamefont {Martis}, \citenamefont {Misra}, \citenamefont {Noy},
  \citenamefont {Pham}, \citenamefont {Qu}, \citenamefont {Rayabharam},
  \citenamefont {Reed}, \citenamefont {Ritt}, \citenamefont {Schwegler},
  \citenamefont {Siwy}, \citenamefont {Strano}, \citenamefont {Wang},
  \citenamefont {Yao}, \citenamefont {Zhan},\ and\ \citenamefont
  {Zhang}}]{Aluru2023}%
  \BibitemOpen
  \bibfield  {author} {\bibinfo {author} {\bibfnamefont {N.~R.}\ \bibnamefont
  {Aluru}}, \bibinfo {author} {\bibfnamefont {F.}~\bibnamefont {Aydin}},
  \bibinfo {author} {\bibfnamefont {M.~Z.}\ \bibnamefont {Bazant}}, \bibinfo
  {author} {\bibfnamefont {D.}~\bibnamefont {Blankschtein}}, \bibinfo {author}
  {\bibfnamefont {A.~H.}\ \bibnamefont {Brozena}}, \bibinfo {author}
  {\bibfnamefont {J.~P.}\ \bibnamefont {de~Souza}}, \bibinfo {author}
  {\bibfnamefont {M.}~\bibnamefont {Elimelech}}, \bibinfo {author}
  {\bibfnamefont {S.}~\bibnamefont {Faucher}}, \bibinfo {author} {\bibfnamefont
  {J.~T.}\ \bibnamefont {Fourkas}}, \bibinfo {author} {\bibfnamefont {V.~B.}\
  \bibnamefont {Koman}}, \bibinfo {author} {\bibfnamefont {M.}~\bibnamefont
  {Kuehne}}, \bibinfo {author} {\bibfnamefont {H.~J.}\ \bibnamefont {Kulik}},
  \bibinfo {author} {\bibfnamefont {H.-K.}\ \bibnamefont {Li}}, \bibinfo
  {author} {\bibfnamefont {Y.}~\bibnamefont {Li}}, \bibinfo {author}
  {\bibfnamefont {Z.}~\bibnamefont {Li}}, \bibinfo {author} {\bibfnamefont
  {A.}~\bibnamefont {Majumdar}}, \bibinfo {author} {\bibfnamefont
  {J.}~\bibnamefont {Martis}}, \bibinfo {author} {\bibfnamefont {R.~P.}\
  \bibnamefont {Misra}}, \bibinfo {author} {\bibfnamefont {A.}~\bibnamefont
  {Noy}}, \bibinfo {author} {\bibfnamefont {T.~A.}\ \bibnamefont {Pham}},
  \bibinfo {author} {\bibfnamefont {H.}~\bibnamefont {Qu}}, \bibinfo {author}
  {\bibfnamefont {A.}~\bibnamefont {Rayabharam}}, \bibinfo {author}
  {\bibfnamefont {M.~A.}\ \bibnamefont {Reed}}, \bibinfo {author}
  {\bibfnamefont {C.~L.}\ \bibnamefont {Ritt}}, \bibinfo {author}
  {\bibfnamefont {E.}~\bibnamefont {Schwegler}}, \bibinfo {author}
  {\bibfnamefont {Z.}~\bibnamefont {Siwy}}, \bibinfo {author} {\bibfnamefont
  {M.~S.}\ \bibnamefont {Strano}}, \bibinfo {author} {\bibfnamefont
  {Y.}~\bibnamefont {Wang}}, \bibinfo {author} {\bibfnamefont {Y.-C.}\
  \bibnamefont {Yao}}, \bibinfo {author} {\bibfnamefont {C.}~\bibnamefont
  {Zhan}}, \ and\ \bibinfo {author} {\bibfnamefont {Z.}~\bibnamefont {Zhang}},\
  }\bibfield  {title} {\enquote {\bibinfo {title} {Fluids and electrolytes
  under confinement in single-digit nanopores},}\ }\href {\doibase
  https://doi.org/10.1021/acs.chemrev.2c00155} {\bibfield  {journal} {\bibinfo
  {journal} {Chem. Rev.}\ }\textbf {\bibinfo {volume} {123}},\ \bibinfo {pages}
  {2737} (\bibinfo {year} {2023})}\BibitemShut {NoStop}%
\bibitem [{\citenamefont {Green}(2022)}]{Green2022}%
  \BibitemOpen
  \bibfield  {author} {\bibinfo {author} {\bibfnamefont {Y.}~\bibnamefont
  {Green}},\ }\bibfield  {title} {\enquote {\bibinfo {title} {Effects of
  surface-charge regulation, convection, and slip lengths on the electrical
  conductance of charged nanopores},}\ }\href {\doibase
  10.1103/PhysRevFluids.7.013702} {\bibfield  {journal} {\bibinfo  {journal}
  {Phys. Rev. Fluids}\ }\textbf {\bibinfo {volume} {7}},\ \bibinfo {pages}
  {013702} (\bibinfo {year} {2022})}\BibitemShut {NoStop}%
\bibitem [{\citenamefont {Peters}\ \emph {et~al.}(2016)\citenamefont {Peters},
  \citenamefont {van Roij}, \citenamefont {Bazant},\ and\ \citenamefont
  {Biesheuvel}}]{Peters2016}%
  \BibitemOpen
  \bibfield  {author} {\bibinfo {author} {\bibfnamefont {P.~B.}\ \bibnamefont
  {Peters}}, \bibinfo {author} {\bibfnamefont {R.}~\bibnamefont {van Roij}},
  \bibinfo {author} {\bibfnamefont {M.~Z.}\ \bibnamefont {Bazant}}, \ and\
  \bibinfo {author} {\bibfnamefont {P.~M.}\ \bibnamefont {Biesheuvel}},\
  }\bibfield  {title} {\enquote {\bibinfo {title} {Analysis of electrolyte
  transport through charged nanopores},}\ }\href {\doibase
  10.1103/PhysRevE.93.053108} {\bibfield  {journal} {\bibinfo  {journal} {Phys.
  Rev. E}\ }\textbf {\bibinfo {volume} {93}},\ \bibinfo {pages} {053108}
  (\bibinfo {year} {2016})}\BibitemShut {NoStop}%
\bibitem [{\citenamefont {Guth}, \citenamefont {Vonau},\ and\ \citenamefont
  {Zosel}(2009)}]{Guth2009}%
  \BibitemOpen
  \bibfield  {author} {\bibinfo {author} {\bibfnamefont {U.}~\bibnamefont
  {Guth}}, \bibinfo {author} {\bibfnamefont {W.}~\bibnamefont {Vonau}}, \ and\
  \bibinfo {author} {\bibfnamefont {J.}~\bibnamefont {Zosel}},\ }\bibfield
  {title} {\enquote {\bibinfo {title} {Recent developments in electrochemical
  sensor application and technology—a review},}\ }\href {\doibase
  https://doi.org/10.1088/0957-0233/20/4/042002} {\bibfield  {journal}
  {\bibinfo  {journal} {Measurement Science and Technology}\ }\textbf {\bibinfo
  {volume} {20}},\ \bibinfo {pages} {042002} (\bibinfo {year}
  {2009})}\BibitemShut {NoStop}%
\bibitem [{\citenamefont {Chouhan}\ and\ \citenamefont
  {Liu}(2011)}]{Chouhan2011}%
  \BibitemOpen
  \bibfield  {author} {\bibinfo {author} {\bibfnamefont {N.}~\bibnamefont
  {Chouhan}}\ and\ \bibinfo {author} {\bibfnamefont {R.-S.}\ \bibnamefont
  {Liu}},\ }\href {\doibase https://doi.org/10.1002/9783527639496.ch1} {\emph
  {\bibinfo {title} {Electrochemical Technologies\\ for Energy Storage and
  Conversion}}}\ (\bibinfo  {publisher} {John Wiley \& Sons, Ltd},\ \bibinfo
  {year} {2011})\ Chap.~\bibinfo {chapter} {1}\BibitemShut {NoStop}%
\bibitem [{\citenamefont {Wu}\ \emph {et~al.}(2022)\citenamefont {Wu},
  \citenamefont {Sun}, \citenamefont {Yu}, \citenamefont {Zhu}, \citenamefont
  {Xu}, \citenamefont {Wu},\ and\ \citenamefont {Xu}}]{Wu2022}%
  \BibitemOpen
  \bibfield  {author} {\bibinfo {author} {\bibfnamefont {X.}~\bibnamefont
  {Wu}}, \bibinfo {author} {\bibfnamefont {S.}~\bibnamefont {Sun}}, \bibinfo
  {author} {\bibfnamefont {X.}~\bibnamefont {Yu}}, \bibinfo {author}
  {\bibfnamefont {X.}~\bibnamefont {Zhu}}, \bibinfo {author} {\bibfnamefont
  {M.}~\bibnamefont {Xu}}, \bibinfo {author} {\bibfnamefont {G.}~\bibnamefont
  {Wu}}, \ and\ \bibinfo {author} {\bibfnamefont {J.}~\bibnamefont {Xu}},\
  }\bibfield  {title} {\enquote {\bibinfo {title} {Review on microfluidic
  construction of advanced nanomaterials for high-performance energy storage
  applications},}\ }\href {\doibase
  https://doi.org/10.1021/acs.energyfuels.2c00576} {\bibfield  {journal}
  {\bibinfo  {journal} {Energy \& Fuels}\ }\textbf {\bibinfo {volume} {36}},\
  \bibinfo {pages} {4708} (\bibinfo {year} {2022})}\BibitemShut {NoStop}%
\bibitem [{\citenamefont {Ibrahim}\ \emph {et~al.}(2022)\citenamefont
  {Ibrahim}, \citenamefont {Navarro-Segarra}, \citenamefont {Sadeghi},
  \citenamefont {Sabaté}, \citenamefont {Esquivel},\ and\ \citenamefont
  {Kjeang}}]{Ibrahim2021}%
  \BibitemOpen
  \bibfield  {author} {\bibinfo {author} {\bibfnamefont {O.~A.}\ \bibnamefont
  {Ibrahim}}, \bibinfo {author} {\bibfnamefont {M.}~\bibnamefont
  {Navarro-Segarra}}, \bibinfo {author} {\bibfnamefont {P.}~\bibnamefont
  {Sadeghi}}, \bibinfo {author} {\bibfnamefont {N.}~\bibnamefont {Sabaté}},
  \bibinfo {author} {\bibfnamefont {J.~P.}\ \bibnamefont {Esquivel}}, \ and\
  \bibinfo {author} {\bibfnamefont {E.}~\bibnamefont {Kjeang}},\ }\bibfield
  {title} {\enquote {\bibinfo {title} {Microfluidics for electrochemical energy
  conversion},}\ }\href {\doibase https://doi.org/10.1021/acs.chemrev.1c00499}
  {\bibfield  {journal} {\bibinfo  {journal} {Chem. Rev.}\ }\textbf {\bibinfo
  {volume} {122}},\ \bibinfo {pages} {7236} (\bibinfo {year}
  {2022})}\BibitemShut {NoStop}%
\bibitem [{\citenamefont {Li}\ \emph {et~al.}(2024)\citenamefont {Li},
  \citenamefont {Huang}, \citenamefont {Lin}, \citenamefont {Fan},
  \citenamefont {Sun}, \citenamefont {Zhou}, \citenamefont {Wang},
  \citenamefont {Bei}, \citenamefont {Zheng}, \citenamefont {Xu}, \citenamefont
  {Wang},\ and\ \citenamefont {Ni}}]{LiLi2024}%
  \BibitemOpen
  \bibfield  {author} {\bibinfo {author} {\bibfnamefont {L.}~\bibnamefont
  {Li}}, \bibinfo {author} {\bibfnamefont {H.}~\bibnamefont {Huang}}, \bibinfo
  {author} {\bibfnamefont {X.-M.}\ \bibnamefont {Lin}}, \bibinfo {author}
  {\bibfnamefont {X.}~\bibnamefont {Fan}}, \bibinfo {author} {\bibfnamefont
  {Y.}~\bibnamefont {Sun}}, \bibinfo {author} {\bibfnamefont {W.}~\bibnamefont
  {Zhou}}, \bibinfo {author} {\bibfnamefont {T.}~\bibnamefont {Wang}}, \bibinfo
  {author} {\bibfnamefont {S.}~\bibnamefont {Bei}}, \bibinfo {author}
  {\bibfnamefont {K.}~\bibnamefont {Zheng}}, \bibinfo {author} {\bibfnamefont
  {Q.}~\bibnamefont {Xu}}, \bibinfo {author} {\bibfnamefont {X.}~\bibnamefont
  {Wang}}, \ and\ \bibinfo {author} {\bibfnamefont {M.}~\bibnamefont {Ni}},\
  }\bibfield  {title} {\enquote {\bibinfo {title} {Enhancing the performance of
  paper-based microfluidic fuel cell via optimization of material properties
  and cell structures: A review},}\ }\href {\doibase
  https://doi.org/10.1016/j.enconman.2024.118255} {\bibfield  {journal}
  {\bibinfo  {journal} {Energy Conversion and Management}\ }\textbf {\bibinfo
  {volume} {305}},\ \bibinfo {pages} {118255} (\bibinfo {year}
  {2024})}\BibitemShut {NoStop}%
\bibitem [{\citenamefont {De}\ \emph {et~al.}(2020)\citenamefont {De},
  \citenamefont {Singh}, \citenamefont {Elias}, \citenamefont {Khare},\ and\
  \citenamefont {Basu}}]{DeBiswajit2020}%
  \BibitemOpen
  \bibfield  {author} {\bibinfo {author} {\bibfnamefont {B.~S.}\ \bibnamefont
  {De}}, \bibinfo {author} {\bibfnamefont {A.}~\bibnamefont {Singh}}, \bibinfo
  {author} {\bibfnamefont {A.}~\bibnamefont {Elias}}, \bibinfo {author}
  {\bibfnamefont {N.}~\bibnamefont {Khare}}, \ and\ \bibinfo {author}
  {\bibfnamefont {S.}~\bibnamefont {Basu}},\ }\bibfield  {title} {\enquote
  {\bibinfo {title} {An electrochemical neutralization energy-assisted
  membrane-less microfluidic reactor for water electrolysis},}\ }\href
  {\doibase https://doi.org/10.1039/D0SE01474E} {\bibfield  {journal} {\bibinfo
   {journal} {Sustainable Energy Fuels}\ }\textbf {\bibinfo {volume} {4}},\
  \bibinfo {pages} {6234} (\bibinfo {year} {2020})}\BibitemShut {NoStop}%
\bibitem [{\citenamefont {Shi}\ \emph {et~al.}(2022)\citenamefont {Shi},
  \citenamefont {Feng}, \citenamefont {Liu}, \citenamefont {Zhang},\ and\
  \citenamefont {Li}}]{SHI2022}%
  \BibitemOpen
  \bibfield  {author} {\bibinfo {author} {\bibfnamefont {T.}~\bibnamefont
  {Shi}}, \bibinfo {author} {\bibfnamefont {H.}~\bibnamefont {Feng}}, \bibinfo
  {author} {\bibfnamefont {D.}~\bibnamefont {Liu}}, \bibinfo {author}
  {\bibfnamefont {Y.}~\bibnamefont {Zhang}}, \ and\ \bibinfo {author}
  {\bibfnamefont {Q.}~\bibnamefont {Li}},\ }\bibfield  {title} {\enquote
  {\bibinfo {title} {High-performance microfluidic electrochemical reactor for
  efficient hydrogen evolution},}\ }\href {\doibase
  https://doi.org/10.1016/j.apenergy.2022.119887} {\bibfield  {journal}
  {\bibinfo  {journal} {Applied Energy}\ }\textbf {\bibinfo {volume} {325}},\
  \bibinfo {pages} {119887} (\bibinfo {year} {2022})}\BibitemShut {NoStop}%
\bibitem [{\citenamefont {Cheng}\ \emph {et~al.}(2022)\citenamefont {Cheng},
  \citenamefont {Li}, \citenamefont {Meena}, \citenamefont {Cao}, \citenamefont
  {Li}, \citenamefont {Kosgei}, \citenamefont {Cheng}, \citenamefont {Luo},
  \citenamefont {Liu}, \citenamefont {Zhu}, \citenamefont {Liu},\ and\
  \citenamefont {Han}}]{cheng2022}%
  \BibitemOpen
  \bibfield  {author} {\bibinfo {author} {\bibfnamefont {S.~K.~S.}\
  \bibnamefont {Cheng}}, \bibinfo {author} {\bibfnamefont {T.}~\bibnamefont
  {Li}}, \bibinfo {author} {\bibfnamefont {S.~S.}\ \bibnamefont {Meena}},
  \bibinfo {author} {\bibfnamefont {Q.}~\bibnamefont {Cao}}, \bibinfo {author}
  {\bibfnamefont {B.}~\bibnamefont {Li}}, \bibinfo {author} {\bibfnamefont
  {B.~K.}\ \bibnamefont {Kosgei}}, \bibinfo {author} {\bibfnamefont
  {T.}~\bibnamefont {Cheng}}, \bibinfo {author} {\bibfnamefont
  {P.}~\bibnamefont {Luo}}, \bibinfo {author} {\bibfnamefont {Q.}~\bibnamefont
  {Liu}}, \bibinfo {author} {\bibfnamefont {G.}~\bibnamefont {Zhu}}, \bibinfo
  {author} {\bibfnamefont {Q.}~\bibnamefont {Liu}}, \ and\ \bibinfo {author}
  {\bibfnamefont {R.~P.}\ \bibnamefont {Han}},\ }\bibfield  {title} {\enquote
  {\bibinfo {title} {Advances in microfluidic technologies for energy storage
  and release systems},}\ }\href {\doibase
  https://doi.org/10.1002/aesr.202200060} {\bibfield  {journal} {\bibinfo
  {journal} {Advanced Energy and Sustainability Research}\ }\textbf {\bibinfo
  {volume} {3}},\ \bibinfo {pages} {2200060} (\bibinfo {year}
  {2022})}\BibitemShut {NoStop}%
\bibitem [{\citenamefont {Nguyen}, \citenamefont {Wereley},\ and\ \citenamefont
  {Shaegh}(2019)}]{nguyen2019fundamentals}%
  \BibitemOpen
  \bibfield  {author} {\bibinfo {author} {\bibfnamefont {N.}~\bibnamefont
  {Nguyen}}, \bibinfo {author} {\bibfnamefont {S.}~\bibnamefont {Wereley}}, \
  and\ \bibinfo {author} {\bibfnamefont {S.}~\bibnamefont {Shaegh}},\
  }\href@noop {} {\emph {\bibinfo {title} {Fundamentals and Applications of
  Microfluidics}}}\ (\bibinfo  {publisher} {Artech House Publishers},\ \bibinfo
  {year} {2019})\BibitemShut {NoStop}%
\bibitem [{\citenamefont {Sparreboom}, \citenamefont {van~den Berg},\ and\
  \citenamefont {Eijkel}(2009)}]{Sparreboom2009}%
  \BibitemOpen
  \bibfield  {author} {\bibinfo {author} {\bibfnamefont {W.}~\bibnamefont
  {Sparreboom}}, \bibinfo {author} {\bibfnamefont {A.}~\bibnamefont {van~den
  Berg}}, \ and\ \bibinfo {author} {\bibfnamefont {J.}~\bibnamefont {Eijkel}},\
  }\bibfield  {title} {\enquote {\bibinfo {title} {Principles and applications
  of nanofluidic transport.}}\ }\href {\doibase 10.1038/nnano.2009.332}
  {\bibfield  {journal} {\bibinfo  {journal} {Nature Nanotech}\ }\textbf
  {\bibinfo {volume} {4}},\ \bibinfo {pages} {713} (\bibinfo {year}
  {2009})}\BibitemShut {NoStop}%
\bibitem [{\citenamefont {Nasiri}\ \emph {et~al.}(2020)\citenamefont {Nasiri},
  \citenamefont {Shamloo}, \citenamefont {Ahadian}, \citenamefont {Amirifar},
  \citenamefont {Akbari}, \citenamefont {Goudie}, \citenamefont {Lee},
  \citenamefont {Ashammakhi}, \citenamefont {Dokmeci}, \citenamefont
  {Di~Carlo},\ and\ \citenamefont {Khademhosseini}}]{Nasiri2020}%
  \BibitemOpen
  \bibfield  {author} {\bibinfo {author} {\bibfnamefont {R.}~\bibnamefont
  {Nasiri}}, \bibinfo {author} {\bibfnamefont {A.}~\bibnamefont {Shamloo}},
  \bibinfo {author} {\bibfnamefont {S.}~\bibnamefont {Ahadian}}, \bibinfo
  {author} {\bibfnamefont {L.}~\bibnamefont {Amirifar}}, \bibinfo {author}
  {\bibfnamefont {J.}~\bibnamefont {Akbari}}, \bibinfo {author} {\bibfnamefont
  {M.~J.}\ \bibnamefont {Goudie}}, \bibinfo {author} {\bibfnamefont
  {K.}~\bibnamefont {Lee}}, \bibinfo {author} {\bibfnamefont {N.}~\bibnamefont
  {Ashammakhi}}, \bibinfo {author} {\bibfnamefont {M.~R.}\ \bibnamefont
  {Dokmeci}}, \bibinfo {author} {\bibfnamefont {D.}~\bibnamefont {Di~Carlo}}, \
  and\ \bibinfo {author} {\bibfnamefont {A.}~\bibnamefont {Khademhosseini}},\
  }\bibfield  {title} {\enquote {\bibinfo {title} {Microfluidic-based
  approaches in targeted cell/particle separation based on physical properties:
  Fundamentals and applications},}\ }\href {\doibase
  https://doi.org/10.1002/smll.202000171} {\bibfield  {journal} {\bibinfo
  {journal} {Small}\ }\textbf {\bibinfo {volume} {16}},\ \bibinfo {pages}
  {2000171} (\bibinfo {year} {2020})}\BibitemShut {NoStop}%
\bibitem [{\citenamefont {Han}\ and\ \citenamefont {Chen}(2020)}]{Han2020}%
  \BibitemOpen
  \bibfield  {author} {\bibinfo {author} {\bibfnamefont {W.}~\bibnamefont
  {Han}}\ and\ \bibinfo {author} {\bibfnamefont {X.}~\bibnamefont {Chen}},\
  }\bibfield  {title} {\enquote {\bibinfo {title} {A review: applications of
  ion transport in micro-nanofluidic systems based on ion concentration
  polarization},}\ }\href {\doibase https://doi.org/10.1002/jctb.6288}
  {\bibfield  {journal} {\bibinfo  {journal} {Journal of Chemical Technology \&
  Biotechnology}\ }\textbf {\bibinfo {volume} {95}},\ \bibinfo {pages} {1622}
  (\bibinfo {year} {2020})}\BibitemShut {NoStop}%
\bibitem [{\citenamefont {Nguyen}\ \emph {et~al.}(2018)\citenamefont {Nguyen},
  \citenamefont {Thach}, \citenamefont {Roy}, \citenamefont {Huynh},\ and\
  \citenamefont {Perrault}}]{mi9090461}%
  \BibitemOpen
  \bibfield  {author} {\bibinfo {author} {\bibfnamefont {H.-T.}\ \bibnamefont
  {Nguyen}}, \bibinfo {author} {\bibfnamefont {H.}~\bibnamefont {Thach}},
  \bibinfo {author} {\bibfnamefont {E.}~\bibnamefont {Roy}}, \bibinfo {author}
  {\bibfnamefont {K.}~\bibnamefont {Huynh}}, \ and\ \bibinfo {author}
  {\bibfnamefont {C.~M.-T.}\ \bibnamefont {Perrault}},\ }\bibfield  {title}
  {\enquote {\bibinfo {title} {Low-cost, accessible fabrication methods for
  microfluidics research in low-resource settings},}\ }\href {\doibase
  https://doi.org/10.3390/mi9090461} {\bibfield  {journal} {\bibinfo  {journal}
  {Micromachines}\ }\textbf {\bibinfo {volume} {9}} (\bibinfo {year} {2018}),\
  https://doi.org/10.3390/mi9090461}\BibitemShut {NoStop}%
\bibitem [{\citenamefont {Werkhoven}\ and\ \citenamefont {van
  Roij}(2020)}]{Werk2020}%
  \BibitemOpen
  \bibfield  {author} {\bibinfo {author} {\bibfnamefont {B.}~\bibnamefont
  {Werkhoven}}\ and\ \bibinfo {author} {\bibnamefont {van Roij}},\ }\bibfield
  {title} {\enquote {\bibinfo {title} {Coupled water, charge and salt transport
  in heterogeneous nano-fluidic systems},}\ }\href@noop {} {\bibfield
  {journal} {\bibinfo  {journal} {Soft Matter}\ }\textbf {\bibinfo {volume}
  {16}},\ \bibinfo {pages} {1527} (\bibinfo {year} {2020})}\BibitemShut
  {NoStop}%
\bibitem [{\citenamefont {Wang}\ \emph {et~al.}(2020)\citenamefont {Wang},
  \citenamefont {Yang}, \citenamefont {Wu}, \citenamefont {Min}, \citenamefont
  {Chen},\ and\ \citenamefont {Hou}}]{Wang2020}%
  \BibitemOpen
  \bibfield  {author} {\bibinfo {author} {\bibfnamefont {S.}~\bibnamefont
  {Wang}}, \bibinfo {author} {\bibfnamefont {X.}~\bibnamefont {Yang}}, \bibinfo
  {author} {\bibfnamefont {F.}~\bibnamefont {Wu}}, \bibinfo {author}
  {\bibfnamefont {L.}~\bibnamefont {Min}}, \bibinfo {author} {\bibfnamefont
  {X.}~\bibnamefont {Chen}}, \ and\ \bibinfo {author} {\bibfnamefont
  {X.}~\bibnamefont {Hou}},\ }\bibfield  {title} {\enquote {\bibinfo {title}
  {Inner surface design of functional microchannels for microscale flow
  control},}\ }\href {\doibase https://doi.org/10.1002/smll.201905318}
  {\bibfield  {journal} {\bibinfo  {journal} {Small}\ }\textbf {\bibinfo
  {volume} {16}},\ \bibinfo {pages} {1905318} (\bibinfo {year}
  {2020})}\BibitemShut {NoStop}%
\bibitem [{\citenamefont {Sinha~Mahapatra}\ \emph {et~al.}(2022)\citenamefont
  {Sinha~Mahapatra}, \citenamefont {Ganguly}, \citenamefont {Ghosh},
  \citenamefont {Chatterjee}, \citenamefont {Lowrey}, \citenamefont {Sommers},\
  and\ \citenamefont {Megaridis}}]{Mahapatra22}%
  \BibitemOpen
  \bibfield  {author} {\bibinfo {author} {\bibfnamefont {P.}~\bibnamefont
  {Sinha~Mahapatra}}, \bibinfo {author} {\bibfnamefont {R.}~\bibnamefont
  {Ganguly}}, \bibinfo {author} {\bibfnamefont {A.}~\bibnamefont {Ghosh}},
  \bibinfo {author} {\bibfnamefont {S.}~\bibnamefont {Chatterjee}}, \bibinfo
  {author} {\bibfnamefont {S.}~\bibnamefont {Lowrey}}, \bibinfo {author}
  {\bibfnamefont {A.~D.}\ \bibnamefont {Sommers}}, \ and\ \bibinfo {author}
  {\bibfnamefont {C.~M.}\ \bibnamefont {Megaridis}},\ }\bibfield  {title}
  {\enquote {\bibinfo {title} {Patterning wettability for open-surface fluidic
  manipulation: Fundamentals and applications},}\ }\href {\doibase
  10.1021/acs.chemrev.2c00045} {\bibfield  {journal} {\bibinfo  {journal}
  {Chemical Reviews}\ }\textbf {\bibinfo {volume} {122}},\ \bibinfo {pages}
  {16752--16801} (\bibinfo {year} {2022})}\BibitemShut {NoStop}%
\bibitem [{\citenamefont {Zeng}, \citenamefont {Wang},\ and\ \citenamefont
  {Guo}(2024)}]{Zeng2024}%
  \BibitemOpen
  \bibfield  {author} {\bibinfo {author} {\bibfnamefont {Q.}~\bibnamefont
  {Zeng}}, \bibinfo {author} {\bibfnamefont {B.}~\bibnamefont {Wang}}, \ and\
  \bibinfo {author} {\bibfnamefont {Z.}~\bibnamefont {Guo}},\ }\bibfield
  {title} {\enquote {\bibinfo {title} {Recent advances in microfluidics by
  tuning wetting behaviors},}\ }\href {\doibase
  https://doi.org/10.1016/j.mtphys.2023.101324} {\bibfield  {journal} {\bibinfo
   {journal} {Materials Today Physics}\ }\textbf {\bibinfo {volume} {40}},\
  \bibinfo {pages} {101324} (\bibinfo {year} {2024})}\BibitemShut {NoStop}%
\bibitem [{\citenamefont {Ge}\ \emph {et~al.}(2020)\citenamefont {Ge},
  \citenamefont {Wang}, \citenamefont {Zhang},\ and\ \citenamefont
  {Yang}}]{Ge2020}%
  \BibitemOpen
  \bibfield  {author} {\bibinfo {author} {\bibfnamefont {P.}~\bibnamefont
  {Ge}}, \bibinfo {author} {\bibfnamefont {S.}~\bibnamefont {Wang}}, \bibinfo
  {author} {\bibfnamefont {J.}~\bibnamefont {Zhang}}, \ and\ \bibinfo {author}
  {\bibfnamefont {B.}~\bibnamefont {Yang}},\ }\bibfield  {title} {\enquote
  {\bibinfo {title} {Micro-/nanostructures meet anisotropic wetting: from
  preparation methods to applications},}\ }\href {\doibase 10.1039/D0MH00768D}
  {\bibfield  {journal} {\bibinfo  {journal} {Mater. Horiz.}\ }\textbf
  {\bibinfo {volume} {7}},\ \bibinfo {pages} {2566} (\bibinfo {year}
  {2020})}\BibitemShut {NoStop}%
\bibitem [{\citenamefont {Ajdari}\ and\ \citenamefont
  {Bocquet}(2006)}]{Adjary2006}%
  \BibitemOpen
  \bibfield  {author} {\bibinfo {author} {\bibfnamefont {A.}~\bibnamefont
  {Ajdari}}\ and\ \bibinfo {author} {\bibfnamefont {L.}~\bibnamefont
  {Bocquet}},\ }\bibfield  {title} {\enquote {\bibinfo {title} {Giant
  amplification of interfacially driven transport by hydrodynamic slip:
  Diffusio-osmosis and beyond},}\ }\href {\doibase
  10.1103/PhysRevLett.96.186102} {\bibfield  {journal} {\bibinfo  {journal}
  {Phys. Rev. Lett.}\ }\textbf {\bibinfo {volume} {96}},\ \bibinfo {pages}
  {186102} (\bibinfo {year} {2006})}\BibitemShut {NoStop}%
\bibitem [{\citenamefont {Masliyah}\ and\ \citenamefont
  {Bhattacharjee}(2006)}]{Mesliyah2006}%
  \BibitemOpen
  \bibfield  {author} {\bibinfo {author} {\bibfnamefont {J.~H.}\ \bibnamefont
  {Masliyah}}\ and\ \bibinfo {author} {\bibfnamefont {S.}~\bibnamefont
  {Bhattacharjee}},\ }\href@noop {} {\emph {\bibinfo {title} {Electrokinetic
  and colloid transport phenomena}}}\ (\bibinfo  {publisher} {John Wiley \&
  Sons},\ \bibinfo {year} {2006})\BibitemShut {NoStop}%
\bibitem [{\citenamefont {Novotný}\ and\ \citenamefont
  {Foret}(2017)}]{Novotny2017}%
  \BibitemOpen
  \bibfield  {author} {\bibinfo {author} {\bibfnamefont {J.}~\bibnamefont
  {Novotný}}\ and\ \bibinfo {author} {\bibfnamefont {F.}~\bibnamefont
  {Foret}},\ }\bibfield  {title} {\enquote {\bibinfo {title} {Fluid
  manipulation on the micro-scale: Basics of fluid behavior in
  microfluidics},}\ }\href {\doibase https://doi.org/10.1002/jssc.201600905}
  {\bibfield  {journal} {\bibinfo  {journal} {Journal of Separation Science}\
  }\textbf {\bibinfo {volume} {40}},\ \bibinfo {pages} {383} (\bibinfo {year}
  {2017})}\BibitemShut {NoStop}%
\bibitem [{\citenamefont {Neto}\ \emph {et~al.}(2005)\citenamefont {Neto},
  \citenamefont {Evans}, \citenamefont {Bonaccurso}, \citenamefont {Butt},\
  and\ \citenamefont {Craig}}]{neto2005}%
  \BibitemOpen
  \bibfield  {author} {\bibinfo {author} {\bibfnamefont {C.}~\bibnamefont
  {Neto}}, \bibinfo {author} {\bibfnamefont {D.~R.}\ \bibnamefont {Evans}},
  \bibinfo {author} {\bibfnamefont {E.}~\bibnamefont {Bonaccurso}}, \bibinfo
  {author} {\bibfnamefont {H.-J.}\ \bibnamefont {Butt}}, \ and\ \bibinfo
  {author} {\bibfnamefont {V.~S.}\ \bibnamefont {Craig}},\ }\bibfield  {title}
  {\enquote {\bibinfo {title} {Boundary slip in newtonian liquids: a review of
  experimental studies},}\ }\href {\doibase 10.1088/0034-4885/68/12/R05}
  {\bibfield  {journal} {\bibinfo  {journal} {Reports on progress in physics}\
  }\textbf {\bibinfo {volume} {68}},\ \bibinfo {pages} {2859} (\bibinfo {year}
  {2005})}\BibitemShut {NoStop}%
\bibitem [{\citenamefont {Wu}\ \emph {et~al.}(2017)\citenamefont {Wu},
  \citenamefont {Chen}, \citenamefont {Li}, \citenamefont {Li}, \citenamefont
  {Xu},\ and\ \citenamefont {Dong}}]{Wu2017}%
  \BibitemOpen
  \bibfield  {author} {\bibinfo {author} {\bibfnamefont {K.}~\bibnamefont
  {Wu}}, \bibinfo {author} {\bibfnamefont {Z.}~\bibnamefont {Chen}}, \bibinfo
  {author} {\bibfnamefont {J.}~\bibnamefont {Li}}, \bibinfo {author}
  {\bibfnamefont {X.}~\bibnamefont {Li}}, \bibinfo {author} {\bibfnamefont
  {J.}~\bibnamefont {Xu}}, \ and\ \bibinfo {author} {\bibfnamefont
  {X.}~\bibnamefont {Dong}},\ }\bibfield  {title} {\enquote {\bibinfo {title}
  {Wettability effect on nanoconfined water flow},}\ }\href {\doibase
  https://doi.org/10.1073/pnas.1612608114} {\bibfield  {journal} {\bibinfo
  {journal} {Proceedings of the National Academy of Sciences}\ }\textbf
  {\bibinfo {volume} {114}},\ \bibinfo {pages} {3358} (\bibinfo {year}
  {2017})}\BibitemShut {NoStop}%
\bibitem [{\citenamefont {Kunert}, \citenamefont {Harting},\ and\ \citenamefont
  {Vinogradova}(2010)}]{Kunert2010}%
  \BibitemOpen
  \bibfield  {author} {\bibinfo {author} {\bibfnamefont {C.}~\bibnamefont
  {Kunert}}, \bibinfo {author} {\bibfnamefont {J.}~\bibnamefont {Harting}}, \
  and\ \bibinfo {author} {\bibfnamefont {O.~I.}\ \bibnamefont {Vinogradova}},\
  }\bibfield  {title} {\enquote {\bibinfo {title} {Random-roughness
  hydrodynamic boundary conditions},}\ }\href {\doibase
  10.1103/PhysRevLett.105.016001} {\bibfield  {journal} {\bibinfo  {journal}
  {Phys. Rev. Lett.}\ }\textbf {\bibinfo {volume} {105}},\ \bibinfo {pages}
  {016001} (\bibinfo {year} {2010})}\BibitemShut {NoStop}%
\bibitem [{\citenamefont {Schmieschek}\ \emph {et~al.}(2012)\citenamefont
  {Schmieschek}, \citenamefont {Belyaev}, \citenamefont {Harting},\ and\
  \citenamefont {Vinogradova}}]{Schmiescheck2012}%
  \BibitemOpen
  \bibfield  {author} {\bibinfo {author} {\bibfnamefont {S.}~\bibnamefont
  {Schmieschek}}, \bibinfo {author} {\bibfnamefont {A.~V.}\ \bibnamefont
  {Belyaev}}, \bibinfo {author} {\bibfnamefont {J.}~\bibnamefont {Harting}}, \
  and\ \bibinfo {author} {\bibfnamefont {O.~I.}\ \bibnamefont {Vinogradova}},\
  }\bibfield  {title} {\enquote {\bibinfo {title} {Tensorial slip of
  superhydrophobic channels},}\ }\href {\doibase 10.1103/PhysRevE.85.016324}
  {\bibfield  {journal} {\bibinfo  {journal} {Phys. Rev. E}\ }\textbf {\bibinfo
  {volume} {85}},\ \bibinfo {pages} {016324} (\bibinfo {year}
  {2012})}\BibitemShut {NoStop}%
\bibitem [{\citenamefont {Garcia-Gonzalez}\ and\ \citenamefont
  {Crick}(2023)}]{Crick23}%
  \BibitemOpen
  \bibfield  {author} {\bibinfo {author} {\bibfnamefont {R.~I.}\ \bibnamefont
  {Garcia-Gonzalez}}\ and\ \bibinfo {author} {\bibfnamefont {C.~R.}\
  \bibnamefont {Crick}},\ }\bibfield  {title} {\enquote {\bibinfo {title} {The
  effect of micro/nano roughness on antifouling and bactericidal surfaces},}\
  }in\ \href {\doibase https://doi.org/10.5772/intechopen.1002808} {\emph
  {\bibinfo {booktitle} {Superhydrophobic Coating}}},\ \bibinfo {editor}
  {edited by\ \bibinfo {editor} {\bibfnamefont {J.}~\bibnamefont {Ou}}}\
  (\bibinfo  {publisher} {IntechOpen},\ \bibinfo {address} {Rijeka},\ \bibinfo
  {year} {2023})\ Chap.~\bibinfo {chapter} {4}\BibitemShut {NoStop}%
\bibitem [{\citenamefont {Rasitha}\ \emph {et~al.}(2024)\citenamefont
  {Rasitha}, \citenamefont {Krishna}, \citenamefont {Anandkumar}, \citenamefont
  {Vanithakumari},\ and\ \citenamefont {Philip}}]{Rasitha2024}%
  \BibitemOpen
  \bibfield  {author} {\bibinfo {author} {\bibfnamefont {T.}~\bibnamefont
  {Rasitha}}, \bibinfo {author} {\bibfnamefont {N.~G.}\ \bibnamefont
  {Krishna}}, \bibinfo {author} {\bibfnamefont {B.}~\bibnamefont {Anandkumar}},
  \bibinfo {author} {\bibfnamefont {S.}~\bibnamefont {Vanithakumari}}, \ and\
  \bibinfo {author} {\bibfnamefont {J.}~\bibnamefont {Philip}},\ }\bibfield
  {title} {\enquote {\bibinfo {title} {A comprehensive review on
  anticorrosive/antifouling superhydrophobic coatings: Fabrication, assessment,
  applications, challenges and future perspectives},}\ }\href {\doibase
  https://doi.org/10.1016/j.cis.2024.103090} {\bibfield  {journal} {\bibinfo
  {journal} {Advances in Colloid and Interface Science}\ }\textbf {\bibinfo
  {volume} {324}},\ \bibinfo {pages} {103090} (\bibinfo {year}
  {2024})}\BibitemShut {NoStop}%
\bibitem [{\citenamefont {Villa}, \citenamefont {Ayuso},\ and\ \citenamefont
  {Castaño-Álvarez}(2019)}]{FERNANDEZLAVILLA2019175}%
  \BibitemOpen
  \bibfield  {author} {\bibinfo {author} {\bibfnamefont {A.~F.~L.}\
  \bibnamefont {Villa}}, \bibinfo {author} {\bibfnamefont {D.~F.~P.}\
  \bibnamefont {Ayuso}}, \ and\ \bibinfo {author} {\bibfnamefont
  {M.}~\bibnamefont {Castaño-Álvarez}},\ }\bibfield  {title} {\enquote
  {\bibinfo {title} {Microfluidics and electrochemistry: an emerging tandem for
  next-generation analytical microsystems},}\ }\href {\doibase
  https://doi.org/10.1016/j.coelec.2019.05.014} {\bibfield  {journal} {\bibinfo
   {journal} {Current Opinion in Electrochemistry}\ }\textbf {\bibinfo {volume}
  {15}},\ \bibinfo {pages} {175} (\bibinfo {year} {2019})},\ \bibinfo {note}
  {innovative Methods in Electrochemistry, Organic and Molecular
  Electrochemistry}\BibitemShut {NoStop}%
\bibitem [{\citenamefont {Malgaretti}\ \emph {et~al.}(2019)\citenamefont
  {Malgaretti}, \citenamefont {Janssen}, \citenamefont {Pagonabarraga},\ and\
  \citenamefont {Rubi}}]{PM19}%
  \BibitemOpen
  \bibfield  {author} {\bibinfo {author} {\bibfnamefont {P.}~\bibnamefont
  {Malgaretti}}, \bibinfo {author} {\bibfnamefont {M.}~\bibnamefont {Janssen}},
  \bibinfo {author} {\bibfnamefont {I.}~\bibnamefont {Pagonabarraga}}, \ and\
  \bibinfo {author} {\bibfnamefont {J.~M.}\ \bibnamefont {Rubi}},\ }\bibfield
  {title} {\enquote {\bibinfo {title} {{Driving an electrolyte through a
  corrugated nanopore}},}\ }\href {\doibase https://doi.org/10.1063/1.5110349}
  {\bibfield  {journal} {\bibinfo  {journal} {The Journal of Chemical Physics}\
  }\textbf {\bibinfo {volume} {151}} (\bibinfo {year} {2019}),\
  https://doi.org/10.1063/1.5110349}\BibitemShut {NoStop}%
\bibitem [{\citenamefont {Vinogradova}, \citenamefont {Silkina},\ and\
  \citenamefont {Asmolov}(2021)}]{Vinogradova2021}%
  \BibitemOpen
  \bibfield  {author} {\bibinfo {author} {\bibfnamefont {O.~I.}\ \bibnamefont
  {Vinogradova}}, \bibinfo {author} {\bibfnamefont {E.~F.}\ \bibnamefont
  {Silkina}}, \ and\ \bibinfo {author} {\bibfnamefont {E.~S.}\ \bibnamefont
  {Asmolov}},\ }\bibfield  {title} {\enquote {\bibinfo {title} {Enhanced
  transport of ions by tuning surface properties of the nanochannel},}\ }\href
  {\doibase https://doi.org/10.1103/PhysRevE.104.035107} {\bibfield  {journal}
  {\bibinfo  {journal} {Phys. Rev. E}\ }\textbf {\bibinfo {volume} {104}},\
  \bibinfo {pages} {035107} (\bibinfo {year} {2021})}\BibitemShut {NoStop}%
\bibitem [{\citenamefont {Vinogradova}, \citenamefont {Silkina},\ and\
  \citenamefont {Asmolov}(2022)}]{Vinogradova2022}%
  \BibitemOpen
  \bibfield  {author} {\bibinfo {author} {\bibfnamefont {O.~I.}\ \bibnamefont
  {Vinogradova}}, \bibinfo {author} {\bibfnamefont {E.~F.}\ \bibnamefont
  {Silkina}}, \ and\ \bibinfo {author} {\bibfnamefont {E.~S.}\ \bibnamefont
  {Asmolov}},\ }\bibfield  {title} {\enquote {\bibinfo {title} {{Transport of
  ions in hydrophobic nanotubes}},}\ }\href {\doibase 10.1063/5.0131440}
  {\bibfield  {journal} {\bibinfo  {journal} {Physics of Fluids}\ }\textbf
  {\bibinfo {volume} {34}},\ \bibinfo {pages} {122003} (\bibinfo {year}
  {2022})}\BibitemShut {NoStop}%
\bibitem [{\citenamefont {Vinogradova}, \citenamefont {Silkina},\ and\
  \citenamefont {Asmolov}(2023)}]{Vinogradova2023}%
  \BibitemOpen
  \bibfield  {author} {\bibinfo {author} {\bibfnamefont {O.~I.}\ \bibnamefont
  {Vinogradova}}, \bibinfo {author} {\bibfnamefont {E.~F.}\ \bibnamefont
  {Silkina}}, \ and\ \bibinfo {author} {\bibfnamefont {E.~S.}\ \bibnamefont
  {Asmolov}},\ }\bibfield  {title} {\enquote {\bibinfo {title} {Slippery and
  mobile hydrophobic electrokinetics: from single walls to nanochannels},}\
  }\href {\doibase https://doi.org/10.1016/j.cocis.2023.101742} {\bibfield
  {journal} {\bibinfo  {journal} {Current Opinion in Colloid \& Interface
  Science}\ ,\ \bibinfo {pages} {101742}} (\bibinfo {year} {2023})}\BibitemShut
  {NoStop}%
\bibitem [{\citenamefont {Cho}, \citenamefont {Chen},\ and\ \citenamefont
  {Chen}(2012)}]{CHO201294}%
  \BibitemOpen
  \bibfield  {author} {\bibinfo {author} {\bibfnamefont {C.-C.}\ \bibnamefont
  {Cho}}, \bibinfo {author} {\bibfnamefont {C.-L.}\ \bibnamefont {Chen}}, \
  and\ \bibinfo {author} {\bibfnamefont {C.-K.}\ \bibnamefont {Chen}},\
  }\bibfield  {title} {\enquote {\bibinfo {title} {Characteristics of combined
  electroosmotic flow and pressure-driven flow in microchannels with
  complex-wavy surfaces},}\ }\href {\doibase
  https://doi.org/10.1016/j.ijthermalsci.2012.06.008} {\bibfield  {journal}
  {\bibinfo  {journal} {International Journal of Thermal Sciences}\ }\textbf
  {\bibinfo {volume} {61}},\ \bibinfo {pages} {94} (\bibinfo {year}
  {2012})}\BibitemShut {NoStop}%
\bibitem [{\citenamefont {Banerjee}\ and\ \citenamefont
  {Nayak}(2019)}]{BANERJEE201917}%
  \BibitemOpen
  \bibfield  {author} {\bibinfo {author} {\bibfnamefont {A.}~\bibnamefont
  {Banerjee}}\ and\ \bibinfo {author} {\bibfnamefont {A.}~\bibnamefont
  {Nayak}},\ }\bibfield  {title} {\enquote {\bibinfo {title} {Influence of
  varying zeta potential on non-newtonian flow mixing in a wavy patterned
  microchannel},}\ }\href {\doibase
  https://doi.org/10.1016/j.jnnfm.2019.05.007} {\bibfield  {journal} {\bibinfo
  {journal} {Journal of Non-Newtonian Fluid Mechanics}\ }\textbf {\bibinfo
  {volume} {269}},\ \bibinfo {pages} {17} (\bibinfo {year} {2019})}\BibitemShut
  {NoStop}%
\bibitem [{\citenamefont {Joly}\ \emph {et~al.}(2004)\citenamefont {Joly},
  \citenamefont {Ybert}, \citenamefont {Trizac},\ and\ \citenamefont
  {Bocquet}}]{Joly2004}%
  \BibitemOpen
  \bibfield  {author} {\bibinfo {author} {\bibfnamefont {L.}~\bibnamefont
  {Joly}}, \bibinfo {author} {\bibfnamefont {C.}~\bibnamefont {Ybert}},
  \bibinfo {author} {\bibfnamefont {E.}~\bibnamefont {Trizac}}, \ and\ \bibinfo
  {author} {\bibfnamefont {L.}~\bibnamefont {Bocquet}},\ }\bibfield  {title}
  {\enquote {\bibinfo {title} {Hydrodynamics within the electric double layer
  on slipping surfaces},}\ }\href {\doibase 10.1103/PhysRevLett.93.257805}
  {\bibfield  {journal} {\bibinfo  {journal} {Phys. Rev. Lett.}\ }\textbf
  {\bibinfo {volume} {93}},\ \bibinfo {pages} {257805} (\bibinfo {year}
  {2004})}\BibitemShut {NoStop}%
\bibitem [{\citenamefont {Harting}, \citenamefont {Kunert},\ and\ \citenamefont
  {Hyv\"aluoma}(2010)}]{Harting2010}%
  \BibitemOpen
  \bibfield  {author} {\bibinfo {author} {\bibfnamefont {J.}~\bibnamefont
  {Harting}}, \bibinfo {author} {\bibfnamefont {C.}~\bibnamefont {Kunert}}, \
  and\ \bibinfo {author} {\bibfnamefont {J.}~\bibnamefont {Hyv\"aluoma}},\
  }\bibfield  {title} {\enquote {\bibinfo {title} {Simulations of slip flow on
  nanobubble-laden surfaces},}\ }\href {\doibase
  10.1088/0953-8984/23/18/184106} {\bibfield  {journal} {\bibinfo  {journal}
  {Microfluidcs and nanofluidics}\ }\textbf {\bibinfo {volume} {8}},\ \bibinfo
  {pages} {1} (\bibinfo {year} {2010})}\BibitemShut {NoStop}%
\bibitem [{\citenamefont {Zhang}\ \emph {et~al.}(2022)\citenamefont {Zhang},
  \citenamefont {Zhang}, \citenamefont {Yang}, \citenamefont {Yue},\ and\
  \citenamefont {Zhang}}]{Zhang2022}%
  \BibitemOpen
  \bibfield  {author} {\bibinfo {author} {\bibfnamefont {Y.}~\bibnamefont
  {Zhang}}, \bibinfo {author} {\bibfnamefont {Z.}~\bibnamefont {Zhang}},
  \bibinfo {author} {\bibfnamefont {J.}~\bibnamefont {Yang}}, \bibinfo {author}
  {\bibfnamefont {Y.}~\bibnamefont {Yue}}, \ and\ \bibinfo {author}
  {\bibfnamefont {H.}~\bibnamefont {Zhang}},\ }\bibfield  {title} {\enquote
  {\bibinfo {title} {A review of recent advances in superhydrophobic surfaces
  and their applications in drag reduction and heat transfer},}\ }\href
  {\doibase https://www.mdpi.com/2079-4991/12/1/44} {\bibfield  {journal}
  {\bibinfo  {journal} {Nanomaterials}\ }\textbf {\bibinfo {volume} {12}}
  (\bibinfo {year} {2022}),\
  https://www.mdpi.com/2079-4991/12/1/44}\BibitemShut {NoStop}%
\bibitem [{\citenamefont {Ranjan}, \citenamefont {Zou},\ and\ \citenamefont
  {Maroo}(2023)}]{Ranjan2023}%
  \BibitemOpen
  \bibfield  {author} {\bibinfo {author} {\bibfnamefont {D.}~\bibnamefont
  {Ranjan}}, \bibinfo {author} {\bibfnamefont {A.}~\bibnamefont {Zou}}, \ and\
  \bibinfo {author} {\bibfnamefont {S.~C.}\ \bibnamefont {Maroo}},\ }\bibfield
  {title} {\enquote {\bibinfo {title} {Durable and regenerative
  superhydrophobic surface using porous nanochannels},}\ }\href {\doibase
  https://doi.org/10.1016/j.cej.2022.140527} {\bibfield  {journal} {\bibinfo
  {journal} {Chemical Engineering Journal}\ }\textbf {\bibinfo {volume}
  {455}},\ \bibinfo {pages} {140527} (\bibinfo {year} {2023})}\BibitemShut
  {NoStop}%
\bibitem [{\citenamefont {Zwanzig}(1992)}]{Zwanzig}%
  \BibitemOpen
  \bibfield  {author} {\bibinfo {author} {\bibfnamefont {R.}~\bibnamefont
  {Zwanzig}},\ }\bibfield  {title} {\enquote {\bibinfo {title} {Diffusion past
  an entropy barrier},}\ }\href {\doibase doi.org/10.1021/j100189a004}
  {\bibfield  {journal} {\bibinfo  {journal} {The Journal of Physical
  Chemistry}\ }\textbf {\bibinfo {volume} {96}},\ \bibinfo {pages} {3926}
  (\bibinfo {year} {1992})}\BibitemShut {NoStop}%
\bibitem [{\citenamefont {Rubi}(2019)}]{entropic2}%
  \BibitemOpen
  \bibfield  {author} {\bibinfo {author} {\bibfnamefont {J.}~\bibnamefont
  {Rubi}},\ }\bibfield  {title} {\enquote {\bibinfo {title} {Entropic diffusion
  in confined soft-matter and biological systems},}\ }\href@noop {} {\bibfield
  {journal} {\bibinfo  {journal} {EPL}\ }\textbf {\bibinfo {volume} {127}},\
  \bibinfo {pages} {10001} (\bibinfo {year} {2019})}\BibitemShut {NoStop}%
\bibitem [{\citenamefont {Rubi}\ \emph {et~al.}(2017)\citenamefont {Rubi},
  \citenamefont {Lervik}, \citenamefont {Bedeaux},\ and\ \citenamefont
  {Kjelstrup}}]{entropic3}%
  \BibitemOpen
  \bibfield  {author} {\bibinfo {author} {\bibfnamefont {J.~M.}\ \bibnamefont
  {Rubi}}, \bibinfo {author} {\bibfnamefont {A.}~\bibnamefont {Lervik}},
  \bibinfo {author} {\bibfnamefont {D.}~\bibnamefont {Bedeaux}}, \ and\
  \bibinfo {author} {\bibfnamefont {S.}~\bibnamefont {Kjelstrup}},\ }\bibfield
  {title} {\enquote {\bibinfo {title} {Entropy facilitated active transport},}\
  }\href {\doibase https://doi.org10.1063/1.4982799} {\bibfield  {journal}
  {\bibinfo  {journal} {J. Chem. Phys.}\ }\textbf {\bibinfo {volume} {146}},\
  \bibinfo {pages} {185101} (\bibinfo {year} {2017})}\BibitemShut {NoStop}%
\bibitem [{\citenamefont {Malgaretti}, \citenamefont {Pagonabarraga},\ and\
  \citenamefont {Rubi}(2013)}]{MPR2013}%
  \BibitemOpen
  \bibfield  {author} {\bibinfo {author} {\bibfnamefont {P.}~\bibnamefont
  {Malgaretti}}, \bibinfo {author} {\bibfnamefont {I.}~\bibnamefont
  {Pagonabarraga}}, \ and\ \bibinfo {author} {\bibfnamefont {M.}~\bibnamefont
  {Rubi}},\ }\bibfield  {title} {\enquote {\bibinfo {title} {Entropic transport
  in confined media: a challenge for computational studies in biological and
  soft-matter systems},}\ }\href {\doibase
  https://doi.org10.3389/fphy.2013.00021} {\bibfield  {journal} {\bibinfo
  {journal} {Frontiers in Physics}\ }\textbf {\bibinfo {volume} {1}} (\bibinfo
  {year} {2013}),\ https://doi.org10.3389/fphy.2013.00021}\BibitemShut
  {NoStop}%
\bibitem [{\citenamefont {Carusela}\ and\ \citenamefont {Rubi}(2017)}]{CR2017}%
  \BibitemOpen
  \bibfield  {author} {\bibinfo {author} {\bibfnamefont {M.}~\bibnamefont
  {Carusela}}\ and\ \bibinfo {author} {\bibfnamefont {J.}~\bibnamefont
  {Rubi}},\ }\bibfield  {title} {\enquote {\bibinfo {title} {Entropic
  rectification and current inversion in a pulsating channel},}\ }\href@noop {}
  {\bibfield  {journal} {\bibinfo  {journal} {J. Chem. Phys.}\ }\textbf
  {\bibinfo {volume} {146}},\ \bibinfo {pages} {184901} (\bibinfo {year}
  {2017})}\BibitemShut {NoStop}%
\bibitem [{\citenamefont {Carusela}, \citenamefont {Malgaretti},\ and\
  \citenamefont {Rubi}(2021)}]{Carusela2021}%
  \BibitemOpen
  \bibfield  {author} {\bibinfo {author} {\bibfnamefont {M.~F.}\ \bibnamefont
  {Carusela}}, \bibinfo {author} {\bibfnamefont {P.}~\bibnamefont
  {Malgaretti}}, \ and\ \bibinfo {author} {\bibfnamefont {J.~M.}\ \bibnamefont
  {Rubi}},\ }\bibfield  {title} {\enquote {\bibinfo {title} {Antiresonant
  driven systems for particle manipulation},}\ }\href {\doibase
  10.1103/PhysRevE.103.062102} {\bibfield  {journal} {\bibinfo  {journal}
  {Phys. Rev. E}\ }\textbf {\bibinfo {volume} {103}},\ \bibinfo {pages}
  {062102} (\bibinfo {year} {2021})}\BibitemShut {NoStop}%
\bibitem [{\citenamefont {Lairez}, \citenamefont {Clochard},\ and\
  \citenamefont {Wegrowe}(2016)}]{Lairez2016}%
  \BibitemOpen
  \bibfield  {author} {\bibinfo {author} {\bibfnamefont {D.}~\bibnamefont
  {Lairez}}, \bibinfo {author} {\bibfnamefont {M.-C.}\ \bibnamefont
  {Clochard}}, \ and\ \bibinfo {author} {\bibfnamefont {J.-E.}\ \bibnamefont
  {Wegrowe}},\ }\bibfield  {title} {\enquote {\bibinfo {title} {The concept of
  entropic rectifier facing experiments},}\ }\href {\doibase
  https://doi.orgdoi.org/10.1038/srep38966} {\bibfield  {journal} {\bibinfo
  {journal} {Scientific Reports}\ }\textbf {\bibinfo {volume} {6}} (\bibinfo
  {year} {2016}),\ https://doi.orgdoi.org/10.1038/srep38966}\BibitemShut
  {NoStop}%
\bibitem [{\citenamefont {Malgaretti}, \citenamefont {Pagonabarraga},\ and\
  \citenamefont {Rubi}(2014)}]{Malgaretti2014EPJ}%
  \BibitemOpen
  \bibfield  {author} {\bibinfo {author} {\bibfnamefont {P.}~\bibnamefont
  {Malgaretti}}, \bibinfo {author} {\bibfnamefont {I.}~\bibnamefont
  {Pagonabarraga}}, \ and\ \bibinfo {author} {\bibfnamefont {J.}~\bibnamefont
  {Rubi}},\ }\bibfield  {title} {\enquote {\bibinfo {title} {Working under
  confinement},}\ }\href {\doibase DOI: 10.1140/epjst/e2014-02334-4} {\bibfield
   {journal} {\bibinfo  {journal} {European Physics Journal Special Topics}\
  }\textbf {\bibinfo {volume} {223}},\ \bibinfo {pages} {3295} (\bibinfo {year}
  {2014})}\BibitemShut {NoStop}%
\bibitem [{\citenamefont {Carusela}\ and\ \citenamefont {Rubi}(2018)}]{CR2018}%
  \BibitemOpen
  \bibfield  {author} {\bibinfo {author} {\bibfnamefont {M.~F.}\ \bibnamefont
  {Carusela}}\ and\ \bibinfo {author} {\bibfnamefont {J.~M.}\ \bibnamefont
  {Rubi}},\ }\bibfield  {title} {\enquote {\bibinfo {title} {Entropy production
  and rectification efficiency in colloid transport along a pulsating
  channel},}\ }\href {\doibase 10.1088/1361-648x/aac0c0} {\bibfield  {journal}
  {\bibinfo  {journal} {Journal of Physics: Condensed Matter}\ }\textbf
  {\bibinfo {volume} {30}},\ \bibinfo {pages} {244001} (\bibinfo {year}
  {2018})}\BibitemShut {NoStop}%
\bibitem [{\citenamefont {Reguera}\ \emph {et~al.}(2012)\citenamefont
  {Reguera}, \citenamefont {Luque}, \citenamefont {Burada}, \citenamefont
  {Schmid}, \citenamefont {Rub\'{\i}},\ and\ \citenamefont
  {H\"anggi}}]{reguera2012}%
  \BibitemOpen
  \bibfield  {author} {\bibinfo {author} {\bibfnamefont {D.}~\bibnamefont
  {Reguera}}, \bibinfo {author} {\bibfnamefont {A.}~\bibnamefont {Luque}},
  \bibinfo {author} {\bibfnamefont {P.~S.}\ \bibnamefont {Burada}}, \bibinfo
  {author} {\bibfnamefont {G.}~\bibnamefont {Schmid}}, \bibinfo {author}
  {\bibfnamefont {J.~M.}\ \bibnamefont {Rub\'{\i}}}, \ and\ \bibinfo {author}
  {\bibfnamefont {P.}~\bibnamefont {H\"anggi}},\ }\bibfield  {title} {\enquote
  {\bibinfo {title} {Entropic splitter for particle separation},}\ }\href
  {\doibase https://doi.org/10.1103/PhysRevLett.108.020604} {\bibfield
  {journal} {\bibinfo  {journal} {Phys. Rev. Lett.}\ }\textbf {\bibinfo
  {volume} {108}},\ \bibinfo {pages} {020604} (\bibinfo {year}
  {2012})}\BibitemShut {NoStop}%
\bibitem [{\citenamefont {Motz}\ \emph {et~al.}(2014)\citenamefont {Motz},
  \citenamefont {Schmid}, \citenamefont {H{\"a}nggi}, \citenamefont {Reguera},\
  and\ \citenamefont {Rubi}}]{motz2014}%
  \BibitemOpen
  \bibfield  {author} {\bibinfo {author} {\bibfnamefont {T.}~\bibnamefont
  {Motz}}, \bibinfo {author} {\bibfnamefont {G.}~\bibnamefont {Schmid}},
  \bibinfo {author} {\bibfnamefont {P.}~\bibnamefont {H{\"a}nggi}}, \bibinfo
  {author} {\bibfnamefont {D.}~\bibnamefont {Reguera}}, \ and\ \bibinfo
  {author} {\bibfnamefont {J.~M.}\ \bibnamefont {Rubi}},\ }\bibfield  {title}
  {\enquote {\bibinfo {title} {Optimizing the performance of the entropic
  splitter for particle separation},}\ }\href {\doibase 10.1063/1.4892615}
  {\bibfield  {journal} {\bibinfo  {journal} {J. Chem. Phys.}\ }\textbf
  {\bibinfo {volume} {141}},\ \bibinfo {pages} {074104} (\bibinfo {year}
  {2014})}\BibitemShut {NoStop}%
\bibitem [{And()}]{Andelman_book}%
  \BibitemOpen
  \href@noop {} {\emph {\bibinfo {title} {Soft condensed matter physics in
  molecular and cell biology}}}\ (\bibinfo  {publisher} {Taylor and Francis},\
  \bibinfo {address} {London})\BibitemShut {NoStop}%
\bibitem [{\citenamefont {Doi}(2013)}]{SoftMat}%
  \BibitemOpen
  \bibfield  {author} {\bibinfo {author} {\bibfnamefont {M.}~\bibnamefont
  {Doi}},\ }\href {\doibase 10.1093/acprof:oso/9780199652952.001.0001} {\emph
  {\bibinfo {title} {Soft Matter Physics}}}\ (\bibinfo  {publisher} {Oxford
  University Press},\ \bibinfo {year} {2013})\BibitemShut {NoStop}%
\end{thebibliography}%

\end{document}